\documentclass[article]{elsarticle}

\usepackage{verbatim}
\usepackage{lineno,hyperref}
\usepackage{subfig}
\usepackage{float}
\usepackage{wrapfig}
\usepackage{todonotes}
\usepackage{tabularx}
\usepackage{amsmath}
\usepackage{amssymb}
\usepackage{graphicx}
\usepackage{times}
\usepackage{url}
\usepackage{multirow}

\usepackage[ruled,vlined,boxed,linesnumbered]{algorithm2e}
\SetKwRepeat{Do}{do}{while}
\SetKwInOut{Input}{input}
\SetKwInOut{Output}{output}

\usepackage{tikz}
\usetikzlibrary{automata,
                arrows,
                calc,
                chains,
                backgrounds,
                datavisualization,
                datavisualization.formats.functions,
                decorations.pathmorphing,
                decorations.pathreplacing,
                decorations.shapes,
                graphs,
                matrix,
                patterns,
                positioning,
                shapes.geometric,
                shapes,
                shapes.geometric,
                shapes.symbols,
                trees
   			  }

\usepackage{pgfplots}
\pgfplotsset{compat=1.10}

\newtheorem{definition}{Definition}
\newtheorem{observation}{Observation}

\graphicspath{{images/}}

\newcommand{\event}{e}
\newcommand{\eventid}{e_{\mid c}}
\newcommand{\eventa}{e_{\mid a}}
\newcommand{\eventime}{e_{\mid t}}

\newcommand{\trace}{\tau}
\newcommand{\traces}{\Pi}
\newcommand{\timestamps}{\mathcal{T}}
\newcommand{\elog}{\mathcal{L}}
\newcommand{\dfg}{\mathcal{G}}
\newcommand{\fact}{\alpha}
\newcommand{\dtz}{\Delta_{0}}
\newcommand{\dtn}{\Delta_{n}}

\begin{document}

\begin{frontmatter}

\title{Process Mining-Driven Analysis of the COVID19 Impact on the Vaccinations of Victorian Patients}

\author{Adriano Augusto\fnref{myfootnote}}
\author{Timothy Deitz}
\author{Noel Faux}
\author{Jo-Anne Manski-Nankervis}
\author{Daniel Capurro}
\address{The University of Melbourne, Australia}
\fntext[myfootnote]{Corresponding author -- a.augusto@unimelb.edu.au}

\begin{abstract}
Process mining is a discipline sitting between data mining and process science, whose goal is to provide theoretical methods and software tools to analyse process execution data, known as event logs. Although process mining was originally conceived to facilitate business process management activities, research studies have shown the benefit of leveraging process mining tools in different contexts, including healthcare. However, applying process mining tools to analyse healthcare process execution data is not straightforward. In this paper, we report the analysis of an event log recording more than 30 million events capturing the general practice healthcare processes of more than one million patients in Victoria--Australia--over five years. Our analysis allowed us to understand benefits and limitations of the state-of-the-art process mining techniques when dealing with highly variable processes and large data-sets. While we provide solutions to the identified limitations, the overarching goal of this study was to detect differences between the patients` health services utilization pattern observed in 2020--during the COVID-19 pandemic and mandatory lock-downs --and the one observed in the prior four years, 2016 to 2019. By using a combination of process mining techniques and traditional data mining, we were able to demonstrate that vaccinations in Victoria did not drop drastically--as other interactions did. On the contrary, we observed a surge of influenza and pneumococcus vaccinations in 2020, contradicting research findings of similar studies conducted in different geographical areas. 
\end{abstract}


\end{frontmatter}


\section{Introduction}\label{sec:intro}

The discipline of Process Mining~\cite{ProcessMiningBook2} was born with the goal to design automated data analysis techniques that could support the phases of the business process management lifecycle~\cite{dumas2013fundamentals}, especially those phases where the
data analysis plays a central role, e.g., process discovery and process monitoring. Over the past two decades, research in the area of process mining has generated a number of \emph{methodologies} and \emph{software tools} (henceforth, process mining techniques). Process mining techniques usually require process execution data, which is known as event logs. 
It is possible to distinguish two major families of process mining techniques~\cite{dumas2013fundamentals}: \emph{operational techniques} and \emph{tactical techniques}. The former
family encompasses techniques whose goal is to generate insights in real-time during the
process execution, e.g., estimating the probability of a negative event to happen; or the likelihood of a specific process outcome.
The latter family encompasses techniques whose goal is to help analysts to discover, analyse,
and periodically monitor the process execution in order to understand how the process
is performed, what are its weaknesses, and how the process can be improved. Two of the most popular tactical process mining techniques, which we will refer throughout this study, are: i) \emph{automated process discovery} -- which allows to automatically discover a process model from event logs; ii) \emph{variant analysis} -- which facilitates the analysis of behavioural differences between process variants (e.g., process instances with a positive outcome versus those with a negative outcome).

Although process mining was initially conceived to be applied within business contexts (such as, banking, wholesaling, manufacturing), research has shown that its value can be harnessed and reused in a multitude of different contexts, including healthcare~\cite{rojas2016process,leonardi2018leveraging,alvarez2018discovering,chen2018data,yang2018approach,martin2020recommendations}. This study sits within the healthcare context, and it is set during the COVID-19 pandemic in Victoria (Australia). Our research goal was to identify differences between the patients` health services utilization pattern observed in 2020--during the COVID-19 pandemic and mandatory lock-downs --and the one observed in the prior four years, 2016 to 2019.
Given that health services are provided via the enacting of healthcare processes, process mining techniques are ideal for achieving our research goal, in particular, process discovery and process variant analysis techniques. To this end, we analysed process execution data extracted from more than 100 general practice (GP) clinics in Victoria. This data included more than 30 million events capturing the GP healthcare processes of more than one million patients in Victoria, over a time-span of approximately five years.

The contributions of this study are on two fronts. 
\begin{itemize}
    \item[--] From a medical perspective, the results of our analysis show that vaccinations in Victoria did not drop as drastically as other clinical interactions did. On the contrary, we observed a surge of influenza and pneumococcus vaccinations in 2020, contradicting research findings of similar studies conducted in different geographical areas in the equivalent seasonal periods~\cite{santoli2020effects,lassi2021impact,Gaythorpe2021}. 
    \item[--] From a process mining perspective, our study highlights the capabilities of state-of-the-art process mining techniques as well as their limitations when dealing with large data-sets recording highly variable processes -- which is typical of the healthcare processes. While we address some of these limitations by providing: i) a method for fixing timestamp equivalence issues in process execution data; and ii) a method to identify boundaries of incomplete traces with unknown start events; we also draw directions for future research in the area of applied process mining in healthcare.
\end{itemize}

The remainder of the paper is structured as follows. In Section~\ref{sec:background} we discuss related work and background. In Section~\ref{sec:analysis}, we describe the data, the analysis we ran, the challenges we faced, and the solution we adopted. In Section~\ref{sec:discussion}, we review the findings of the data analysis, providing a medical interpretation and considering their consequences. Lastly, Section~\ref{sec:conclusion} summarises our results and draws the conclusion.
\section{Background and Related Work}\label{sec:background}

\subsection{Process Mining in Healthcare}

The analysis of healthcare delivery from the process perspective has been a core aspect of health services research and redesign. However, until recently, the analysis of healthcare data using a process perspective has been challenging due to the limited availability of electronic health data (and/or its poor quality) and the lack of powerful methods to quickly make sense of it~\cite{mans2012process,martin2020recommendations}. The recent adoption of electronic health records alongside process-aware information systems~\cite{dumas2005process} has generated vast amounts of healthcare data--both clinical and administrative--that can be leveraged to better understand healthcare processes. Several systematic reviews have highlighted the use and potential benefits of applying process mining methods to understand and improve healthcare processes~\cite{rojas2016process, erdogan2018systematic, batista2018process,martin2020recommendations}, and research reports include uses of a range of process mining techniques such as automated process discovery, conformance checking, and process variant analysis. 

Automated process discovery techniques~\cite{weijters2011flexible,vanden2017fodina,augusto2018split,leemans2014infrequent,augusto2018automated} allow one to discover patients' clinical pathways from the recordings of their healthcare process activities captured by the hospitals and clinics information systems~\cite{mans2008application,mans2012process}. 
Conformance checking techniques~\cite{carmona2018conformance, dunzer2019conformance} allow one to automatically compare the observed healthcare process behaviour (in the form of process execution data) against a prescribed process behaviour to identify differences between actual and normative healthcare behaviour. The latter is usually provided in the form of an imperative process model or as a set of declarative process rules~\cite{rovani2015declarative}, which rather than capturing the full process behaviour may describe clinical guidelines.  
Process variant analysis techniques~\cite{bolt2018process,taymouri2020business,taymouri2021business,cecconi2021detection} allow one to automatically compare two or more sets of healthcare process executions exhibiting different outcomes (or performance) to identify relevant differences between the executions that may have had an impact on the outcome or performance of the healthcare process. These type of techniques are applied to answer questions such as: what were the differences between the healthcare treatments provided by two different hospitals to patients having the same diagnosis? 

One of the earliest application of process mining in healthcare dates back to 2008, Mans et al.~\cite{mans2008application} used Heuristics Miner~\cite{weijters2011flexible} to extract insights from healthcare process data, both from clinical and administrative perspective, including process handovers analysis by levering the process mining analytics platform ProM.~\footnote{\scriptsize{\url{https://www.promtools.org/}}}
Poelmans et al.~\cite{poelmans2010combining} used a combination of process mining and data mining techniques to detect and analyse differences in the healthcare pathways of patients treated for breast cancer and how they would respond to different therapies. 
Lakshmanan et al.~\cite{lakshmanan2013investigating} proposed an approach for discovering patients healthcare pathways and correlate them to their outcomes, combining techniques from process mining and data mining (including clustering and pattern mining).
Suriadi et al.~\cite{suriadi2014measuring} applied process mining techniques to understand the differences of the treatments provided to patients suffering from chest pain at four South Australian hospitals.
Partington et al.~\cite{partington2015process} applied process mining techniques to analyse the quality and the costs of the healthcare services provided to patients at one South Australian hospital.
Roviani et al.~\cite{rovani2015declarative} reported a case study on how to leverage declarative process mining techniques to identify divergences between clinical guidelines and the observed execution of clinical processes, at the urology department of the Isala hospital in the Netherlands.
Leonardi et al.~\cite{leonardi2018leveraging} proposed a method to abstract low-level process execution data (in the form of simple actions), turning it into high-level data that can be used for process mining applications. They validated their method by discovering process models from healthcare services, showing that their method improved the graphical representation of the healthcare processes, and facilitated the clustering of similar process executions.
Alvarez et al.~\cite{alvarez2018discovering} applied process mining techniques to discover process models capturing how healthcare professionals operate within emergency rooms, analysing them to identify opportunities for process improvement.
Chen et al.~\cite{chen2018data} proposed a framework to extract high-level descriptions of medical treatment processes from electronic medical records by applying clustering techniques on doctor order set sequences. Their framework allows to enrich the extracted process descriptions with additional information regarding the process performance (e.g., cost, length), providing support for improvement.   
Yang et al.~\cite{yang2018approach} designed a process mining approach to automatically and in real-time detect process deviations from recommended clinical guidelines. They validated their approach on a set of pediatric trauma resuscitation procedures, demonstrating the effectiveness of their solution.

All these studies on process mining in healthcare represent only a fraction of the existing ones, but reporting on all of them would require a separate study and it would be outside the scope of this one. Hence, we refer the interested reader to the latest literature reviews~\cite{batista2018process,erdogan2018systematic}. 

Given the diversity of tools available and the applicability of process mining to healthcare, we used this perspective to understand changes in health services utilization patterns during the COVID-19 pandemic in Australia.

\subsection{Process changes during the COVID-19 pandemic}

Since the early months of the COVID-19 pandemic, the main drivers behind lock-downs and stay-at-home measures were the need to reduce face-to-face interactions to prevent the virus from spreading uncontrollably, the subsequent increase in morbidity, mortality, and overwhelming of healthcare service providers. 

In parallel, there were growing concerns that stay-at-home recommendations, lockdown measures, and the fear of becoming infected would have a deep impact on the provision of non-COVID-19 health services. Although heterogeneous, most governments across the world recommended some form of mobility reduction measures to reduce the transmission rate of SARS-CoV-19 so the expectations were that most countries would be impacted, although at different extents. Several publications reported the observed effects on the utilization of health services. The World Stroke Organization reported on a reduction on the number of patients being diagnosed with stroke despite COVID-19 apparently increasing the risk of this diseases and attributed the change to reduced access to health services~\cite{markus2020covid}. These findings were confirmed in the USA~\cite{dula2020decrease}. Similar effects were described for patients with acute myocardial infarction~\cite{kulkarni2020covid}, and cancer~\cite{jazieh2020impact, eskander2021access}, among other conditions. This phenomenon was also observed for preventative care services such as cancer screening~\cite{van2021impact,d2021impact, CancerAustralia}, and maternal and child health services~\cite{roberton2020early}. In particular, there were growing concerns that a significant reduction in immunizations would result in an increase in vaccine-preventable conditions~\cite{WHO_2020, santoli2020effects}. 

The goal of this study was to analyse changes in health services utilization patterns during the 2020 COVID-19 pandemic and associated lock-downs in Victoria (Australia).

\section{Analysis, Observations, and Challenges}\label{sec:analysis}

In this section, we introduce the data we analysed, discussing its characteristics and highlighting those that are the most critical in the context of this study. We describe what methodology and tools we used to analyse the data, what findings we uncovered, what challenges we faced during the analysis and how we addressed them. While we were able to solve some of these challenges, by proposing approaches that can be reused in different contexts, other challenges remain open or partially addressed and should be considered in future research work in the area of process mining. 

\subsection{Preliminaries}

Before discussing our analysis, we provide some formal definition for the concepts we refer to throughout section. While we contextualised these definitions within our study, we remark that these are well-known definitions and concepts in the area of process mining~\cite{ProcessMiningBook2}.

\begin{definition}\textbf{Event --}\label{def:event}
An \emph{event} $\event$ captures the execution of an activity within a process instance. An event can be represented as a tuple $(x_1, x_2, \dots, x_n)$, where each element $x_i$ captures an attribute of the event, and at least three attributes are present: the process instance ID ($c$ -- event ID); the label of the activity the event refers to ($a$ -- event activity); and the timestamp ($t$ -- event timestamp). Additional attributes usually capture the process resource who executed the activity, customer information, etc. In the following, given an event $\event$, we will refer to its three required attributes with the notation $\eventid$, $\eventa$, $\eventime$.
\end{definition}

\begin{definition}\textbf{Event Log --}\label{def:log}
An \emph{event log} $\elog$ is a sequence of events $\langle \event_1, \event_2, \dots, \event_n \rangle$, such that all the events are ordered by their timestamp. Formally, $\forall \event_i \in \elog \mid i \in [1, n-1] \cap \mathbb{N} \Rightarrow {\event_i}_{\mid t} \leq {\event_{i+1}}_{\mid t}$.
\end{definition}

\begin{definition}\textbf{Trace\footnote{} --}\label{def:trace}
Given an event log $\elog$, a \emph{trace} of the event log $\trace \in \elog$ is a sequence of events, $\trace = \langle \event_1, \event_2, \dots, \event_n \rangle$, such that all the events belong to the event log, all the events are ordered by their timestamp, and all the events have the same event ID attribute. Formally, $\forall \event_i \in \trace \mid i \in [1, n-1] \cap \mathbb{N} \Rightarrow {\event_i}_{\mid c} = {\event_{i+1}}_{\mid c} \wedge {\event_i}_{\mid t} \leq {\event_{i+1}}_{\mid t}$.
\end{definition}

We note that, according to Definition~\ref{def:trace}, we can also consider an event log as a multiset of traces. 

\begin{definition}\textbf{Directly-follows Relation --}\label{def:dfr}
Given an event log $\elog = \langle \event_1, \event_2, \dots, \event_n \rangle$, we say that a \emph{directly-follows relation} holds between any two events $\event_i, \event_j \in \elog$ if and only if $\event_i$ and $\event_j$ belong to the same trace and $j = i+1$, in other words, the two events follow each other in (at least) one trace. We indicate such a relation with the notation $\event_i \rightarrow \event_j$. Formally, given $\event_i, \event_j \in \elog$, $\event_i \rightarrow \event_j \Longleftrightarrow \exists \trace \in \elog \mid \event_i, \event_j \in \trace \wedge j = i+1$. We extend the concept of directly-follows relation to the event activities, i.e., if  $\event_i \rightarrow \event_j$ then we say that also ${\event_i}_{\mid a} \rightarrow {\event_j}_{\mid a}$ holds.
\end{definition}

\begin{definition}\textbf{Directly-Follows Graph (DFG) --}
Given an event log $\elog$, its \emph{Directly-Follows Graph (DFG)} is a directed graph $\dfg = (N, E)$, where:
$N$ is the set of nodes, $N = \{ n \mid \exists \event \in \elog \wedge \eventa = x \}$;
and $E$ is the set of edges $E = \{(x, y) \in N \times N \mid \exists \event_1, \event_2 \in \elog \wedge \event_1 \rightarrow \event_2 \wedge {\event_1}_{\mid a} =x \wedge {\event_2}_{\mid a} = y \}$. In other words, each node of the DFG represents a unique activity recorded in the event log, and each edge of the DFG represents a directly-follows relation between two activities -- represented by the source node and target node of the edge.
\end{definition}

\begin{definition}\textbf{(Business) Process~\cite{dumas2013fundamentals} --}\label{def:process}
A (Business) Process is a sequence of events, activities, and decisions involving actors and data objects
triggered by a specific start event and leading to a specific end event (i.e., process outcome) that delivers value to a customer.
\end{definition}

\subsection{Dataset}\label{sec:dataset}

In this study we used the \emph{Patron dataset}~\cite{Patron2021}. This dataset stores de-identified patient data from the Patron primary care data repository (extracted from consenting general practices), that has been created and is operated by the Department of General Practice at The University of Melbourne~\cite{Patron2021, canaway2019gathering}. This dataset is aggregated from  more than 100 General Practice (GP) clinics in Victoria (Australia) and includes both administrative and clinical data, including all interactions between patients and their GPs, for more than one million patients. Access to the data was approved by the Melbourne Health Human Research Ethics Committee (HREC). The dataset is stored in a relational database, which includes the following six tables: Patient Details (Demographics);  Patient Clinical Information; Medical History (Diagnoses); Patient Visits; Medications; Investigations (Pathology and Imaging). While the first three tables contain information regarding the patient and their clinical history; the last three tables containing information regarding the patient healthcare processes, respectively: information on patient visits to and interactions with their GP doctor(s); information on patient drugs prescriptions; and information on patient pathology and imaging tests and results.

Looking at the latter three tables through the lens of Definition~\ref{def:event}, an \emph{event ID} corresponds to a patient ID, which identifies a unique patient accessing GP services across all the tables. An \emph{event activity} corresponds to a medical activity the patient underwent. From the three tables capturing the patient healthcare process, it is possible to extract seven medical activities, which are reported in Table~\ref{tab:actmap}. These activities capture all the interactions of a patient with their GP, including the drugs they have been prescribed, their pathology and imagining tests and results, and their vaccinations. For simplicity, we will refer to each of these seven activities by using a letter $A$ to $G$ (following the mapping in Table~\ref{tab:actmap}). Lastly, the \emph{event timestamp} corresponds to the time a medical activity was completed. We note that the Patron dataset does not record information regarding activities' lifecycle, e.g., the start and the completion of the activities, and that the timestamp granularity is at day-level (i.e., the smallest difference between timestamps is at day-level). Such a timestamp granularity is frequent in the healthcare contexts, and (at least in our case) it is related to how the system records events into the database. Consequently, it was virtually impossible to infer a better timestamp granularity (e.g., hours and minutes) or the duration of a single medical activity (e.g., how long a GP visit would last).
In light of this, in the Patron dataset, a \emph{trace} captures a unique patient accessing GP services over the time, i.e., a process instance of the GP day-to-day healthcare process. 
\begin{table}[htbp]
  \centering
  {\scriptsize{
  \caption{Encoding of the healthcare activities}
  \vspace{-4mm}
    \begin{tabular}{r|l}
            \hline
          \textbf{Activity Label} & \textbf{Activity Description}\\\hline
          \textbf{A} & Patient attends a GP doctor visit\\\hline
          \textbf{B} & GP records a measurement (e.g., blood pressure)\\\hline
          \textbf{C} & Patient is prescribed a medication\\\hline
          \textbf{D} & Patient is prescribed a medication refill\\\hline
          \textbf{E} & Patient is referred for a laboratory or imaging study (e.g., blood analysis)\\\hline
          \textbf{F} & Tests results are recorded\\\hline
          \textbf{G} & One or more vaccinations are administered/recorded\\\hline
    \end{tabular}%
  \label{tab:actmap}%
    }}
\end{table}%

The de-identified data was stored in a secure virtual machine. While this was a strict requirement for analysing the data, such a secure environment posed some challenges during the data analysis stage (discussed later in this section), mostly related to the fact that it did not allow for internet access.

\subsection{Methodology}\label{sec:method}

To conduct our analysis, we adhered to the methodology proposed by van Eck et al.~\cite{van2015pm}, adapting it to our context. The PM\textsuperscript{2} methodology~\cite{van2015pm} has six stages: planning; data extraction; data processing; data mining and analysis; evaluation; and process improvement and support. We thoroughly executed all the stages with the exception of the last stage. Given that this study did not involve GP clinics and healthcare practitioners, we did not have the means to implement a redesigned process, besides, it would have been outside the scope of this study.

\subsection{Planning, Data Extraction and Processing}

Following the PM\textsuperscript{2} methodology, we started from the planning, which includes three steps: i) selecting the process to analyse; ii) determining the process analysis goal; iii) and assembling a team. Indirectly, the Patron dataset drove the process selection. Given that it captures ambulatory patients' interactions with their GPs, we selected for our analysis the GP day-to-day healthcare process. Our analysis objective was to identify differences between the GP healthcare services provided in 2020--during the COVID-19 pandemic and mandatory lock-downs --and those observed in the prior four years, 2016 to 2019. The authors of this paper composed the research team, bringing expertise in process mining, data mining, and (medical) general practice. 

Once the scope of our analysis was set, we moved to data extraction and processing. 
The Patron dataset, as mentioned above, already included all the data we required to analyse the selected process. The extraction of this data was performed outside this study, and it is not our contribution. However, healthcare process data rarely comes in the form of ready-to-use event logs~\cite{mans2012process,partington2015process}, which is the required data format for conducting a process mining analysis~\cite{ProcessMiningBook2,van2015pm}, and the Patron dataset was no exception.
During this stage, we focused on transforming the available data into an event log that could allow us to achieve our analysis goal.
This required us to identify what entries of the relational database were suitable to be turned into events. As mentioned above, we extract all the entries from three tables out of six, which captured the medical activities shown in Table~\ref{tab:actmap}. Each entry of a table included the patient ID and the timestamp, hence, the conceptual mapping from table entries to events was straightforward. We note that this mapping was facilitated by the existing of a very extensive data dictionary describing the Patron dataset, which often is not available. 

The data extracted captured a time-span of (almost) five years, from January 2016 to November 2020, but we reduced this time-span to keep only the data collected between 01-March to 30-November for the years 2016, 2017, 2018, 2019, and 2020. This choice was driven by three factors: i) our analysis goal (as mentioned above); ii) a key date in the international response to the COVID-19 pandemic; iii) and our latest access to data. Precisely, given that the World Health Organization (WHO) officially declared the COVID-19 a pandemic on the 11\textsuperscript{th} March, we set the start date of our analysis on the 1\textsuperscript{st} of March, while for the end date we were forced to set it to the 30\textsuperscript{th} of November, which was our latest available access to data. We also note that the two dates are closely related to the enforcement of the first lockdown restrictions in Victoria (16-March-2020) and the lifting of the last lockdown restrictions in Victoria (09-November-2020), in the year 2020.

The data was extracted via an ad-hoc R-script and saved in the form of CSV event logs. These CSV logs were then converted in the standard XES format via Apromore (academic version)~\footnote{\scriptsize{\url{https://apromore.org/academic-alliance/}}}, which can be used without internet access and does not have limits on the amount of data to be processed, as opposed to Disco~\footnote{\scriptsize{\url{https://fluxicon.com/disco/}}} or Celonis~\footnote{\scriptsize{\url{https://www.celonis.com/academic-alliance/}}}. Alternatively, we could have converted the CSV event logs into XES format via ProM~\footnote{\scriptsize{\url{https://www.promtools.org/}}}

Once we obtained the event logs from the Patron dataset, we proceeded to the data mining and analysis stage.

\subsection{Data Analysis and Initial Observations}\label{sec:observations}

By looking at the data through the lens of Definition~\ref{def:event},~\ref{def:log}, and~\ref{def:trace}, we could summarise its characteristics as shown in Table~\ref{tab:logs}.
\begin{table}[htbp]
  \centering
  {\scriptsize{
  \caption{Event logs characteristics}
  \vspace{-4mm}
  \begin{tabular}{r|c|rr|cc|rr|ccc}
    \hline
    \textbf{Event}
    & \multicolumn{3}{c|}{\textbf{Traces}}
    & \multicolumn{4}{c|}{\textbf{Events}}
    & \multicolumn{3}{c}{\textbf{Trace length}}
    \\\cline{2-11}
    \textbf{Log}
    & \textbf{Total}
    & \multicolumn{2}{|c|}{\textbf{Distinct (\#) (\%)}}
    & \textbf{Total}
    & \textbf{Distinct}
    & \multicolumn{2}{|c|}{\textbf{Filtered (\#) (\%)}}
    & \textbf{Min}
    & \textbf{Avg}
    & \textbf{Max} \\\hline
    
    \textbf{GP16-20} & 2,482,587 & 1,028,328 & 41.4 & 31,769,481 & 7 & 1,156,054 & 3.6 & 1 & 12 & 2317 \\
    \textbf{GP20} & 401,370 & 167,107 & 41.6 & 5,106,686 & 7 & 205,213 & 4.0 & 1 & 12 & 2317	\\	
    \textbf{GP19} & 522,022 & 221,789 & 42.5 & 6,868,762 & 7 & 266,904 & 3.9 & 1 & 12 & 1966 \\
    \textbf{GP18} & 531,618 & 221,732 & 41.7 & 6,865,778 & 7 & 229,803 & 3.3 & 1 & 12 & 748 \\
    \textbf{GP17} & 520,502 & 215,079 & 41.3 & 6,696,341 & 7 & 221,548 & 3.3 & 1 & 12 & 834 \\
    \textbf{GP16} & 507,075 & 202,621 & 40.0 & 6,231,914 & 7 & 232,586 & 3.7 & 1 & 11 & 652 \\\hline
    \end{tabular}
    \label{tab:logs}
    }}
\end{table}

The main log (labeled, GP16-20) covered 45 months. The GP16-20 log counts almost 2.5 million traces (short of 20 thousand), of which 1.0 million (41.4\%) are distinct -- meaning no duplicate of that trace is present in the log. These traces include 31.8 million events, which -- to the best of our knowledge -- dwarf any of the real-life public logs used in automated process discovery research~\cite{augusto2018automated}. The trace length varies widely, with minimum, average, and maximum length of 1, 12, and 2317 events (respectively).

Given that our goal was to compare the patients behaviour in the months between March and November 2020 against the patients behaviour in the same timeframe of the past four years, we divided the log into five sublogs (namely, GP20, GP19, GP18, GP17, GP16), each of them capturing the 9-month timeframe in one of the five years under analysis. Such an approach is common for performing process behavioural comparison -- known in the area of process mining as process variant analysis~\cite{taymouri2021business}. Looking at Table~\ref{tab:logs}, we notice that dividing the GP16-20 log into five sublogs does not affect much the variety of the process behaviour. Although the absolute number of events and traces reduces, each of the five (sub)logs maintains remarkable characteristics; i.e., 5.1 million (GP20 log) to 6.9 million (GP19) events, and 401 thousand (GP20 log) to 520 thousand (GP17 log) traces (on average, 41\% distinct). As a comparison, the largest real-life event log used in the series of business process intelligence challenges had 1.6 million events.\footnote{\scriptsize{\url{https://icpmconference.org/2019/icpm-2019/contests-challenges/bpi-challenge-2019/}}} By analysing the characteristics of these five logs, we can immediately draw some initial observations.

\begin{observation}\label{obs1}
In 2020, there was an average drop of 22.8\% of patients accessing GP clinic healthcare services, compared to 2016-19. This is captured by the decrease of the total number of traces observed in the GP20 log, 401,370 as opposed to an average of 520,304 across the previous years -- having min and max of 507,075 and 531,618. 
\end{observation}

\begin{observation}\label{obs2}
In 2020, GP clinic healthcare processes maintained their high-level overarching variety. This is captured by the almost constant percentage of distinct traces, 41.6\% in 2020, and 41.4\% on average over 2016-19.
Meaning that each healthcare process instance was observed exactly the same little more than two times.
\end{observation}

Observation~\ref{obs1} was expected, given that a strict lockdown was enforced in Victoria from 16-March to 21-June and from 04-July to 09-Nov, that possibly deterred patients from accessing healthcare for what they considered minor issues. Observation~\ref{obs2} had a surprising nature, in fact, intuitively, we would expect that the combination of lockdown and pandemic would foster standardization in healthcare processes (i.e., less variability). 

\begin{figure}[H]
	\centering
	\subfloat[Activity absolute frequencies]{
		\includegraphics[width=0.45\textwidth]{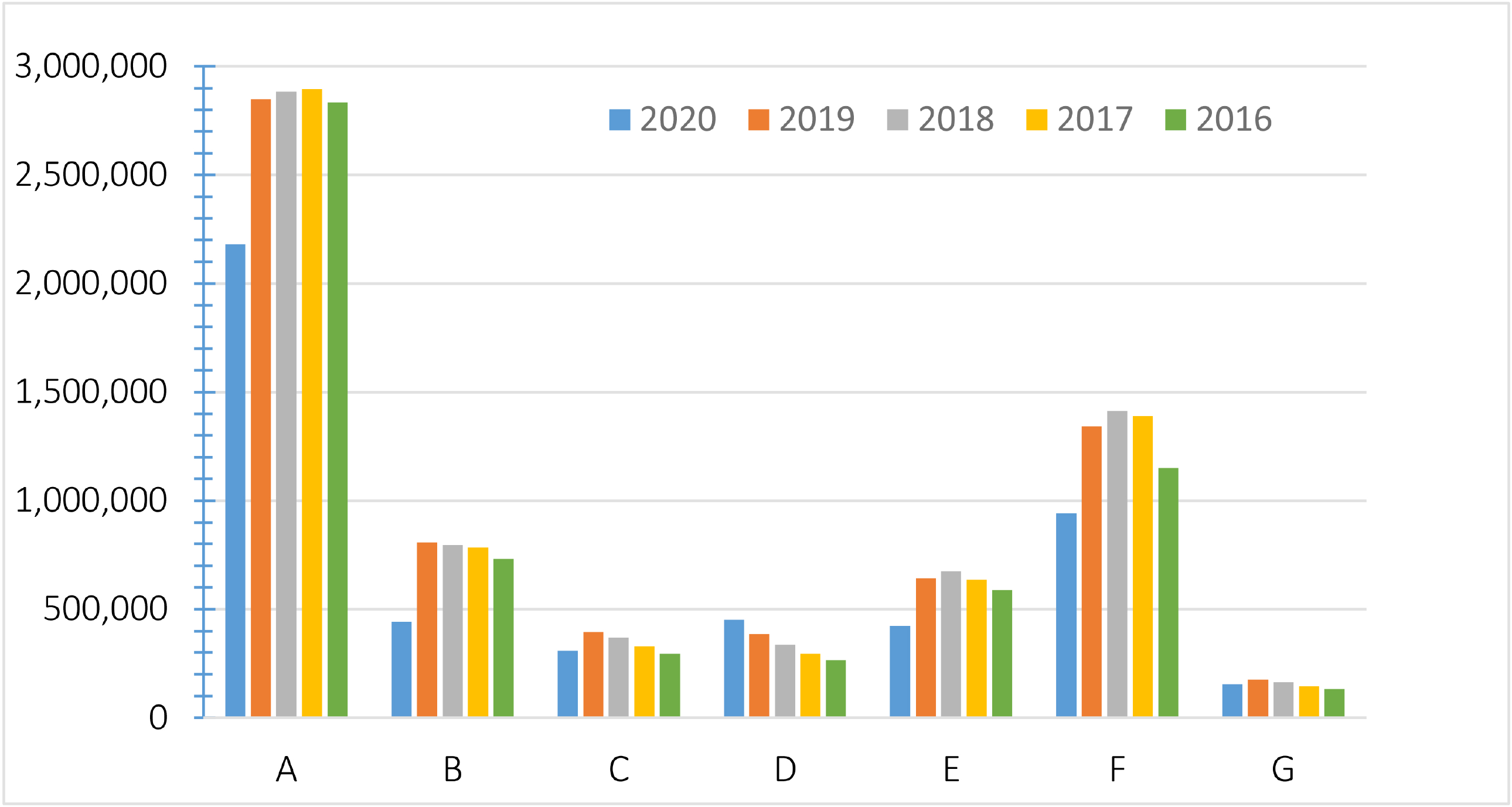}\label{fig:absfreq}
	}
	\hfill
	\subfloat[Activity relative frequencies]{
		\includegraphics[width=0.45\textwidth]{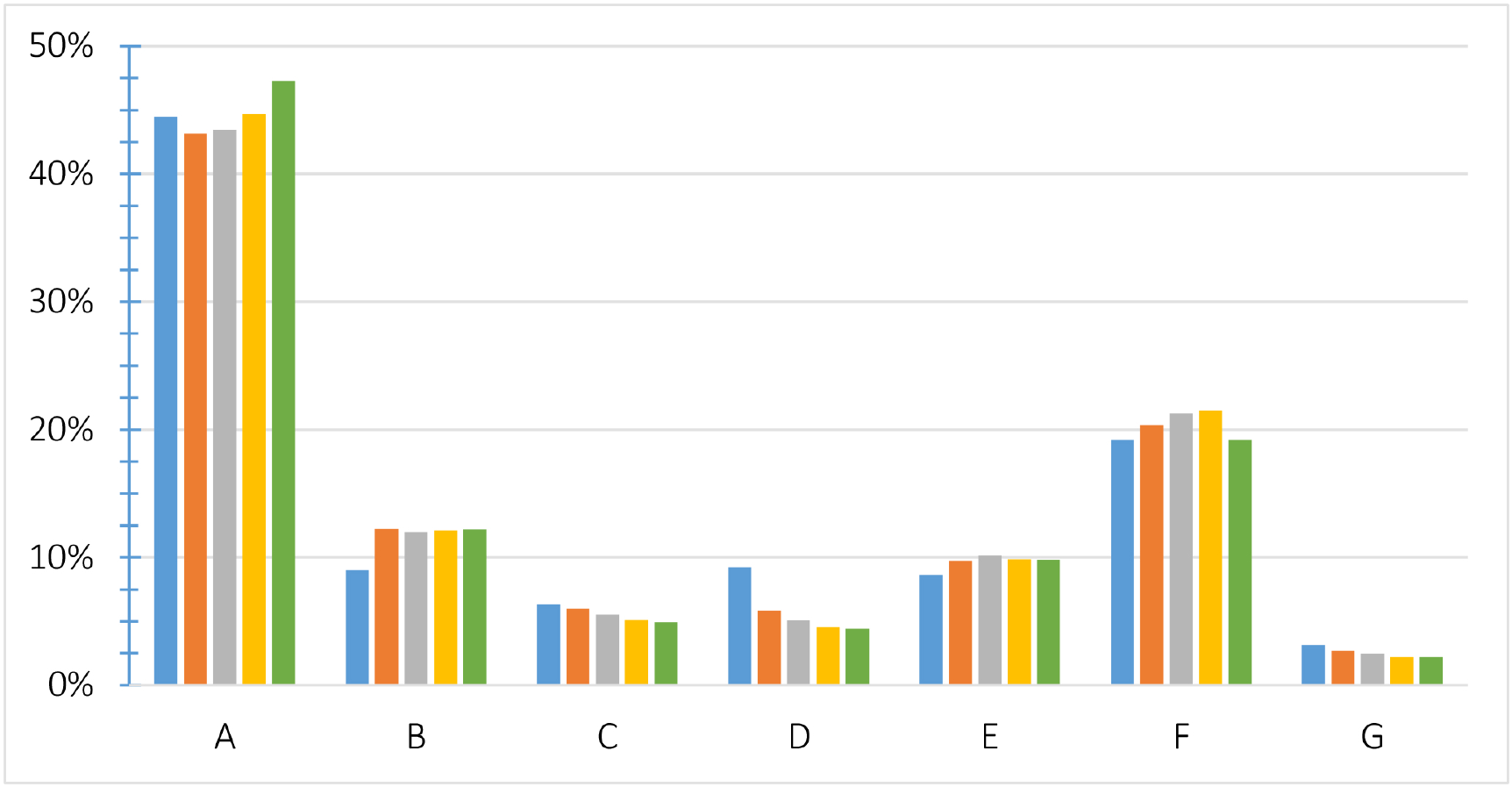}\label{fig:relfreq}
	}
		\\
	\subfloat[Activity A - absolute frequency by month]{
		\includegraphics[width=0.45\textwidth]{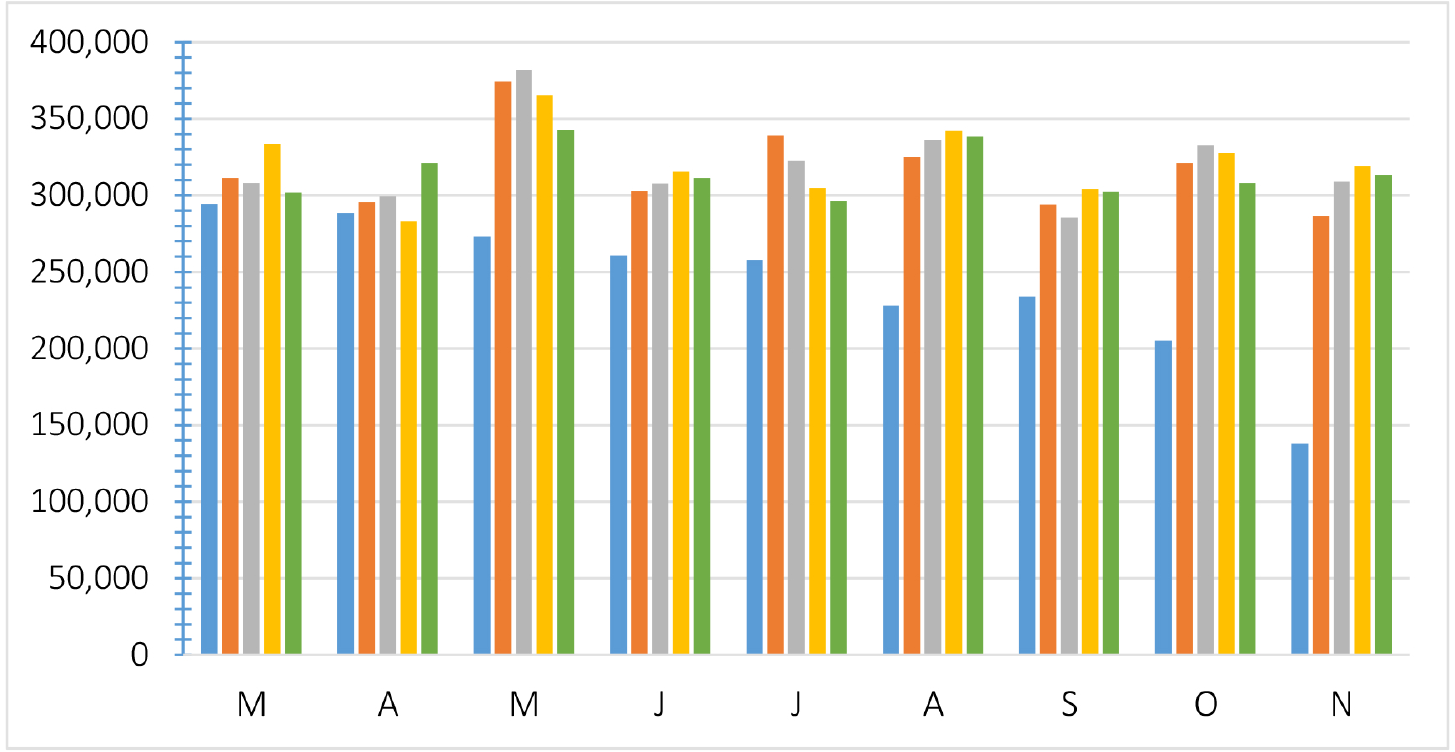}\label{fig:disA}
	}
	\hfill
	\subfloat[Activity B - absolute frequency by month]{
		\includegraphics[width=0.45\textwidth]{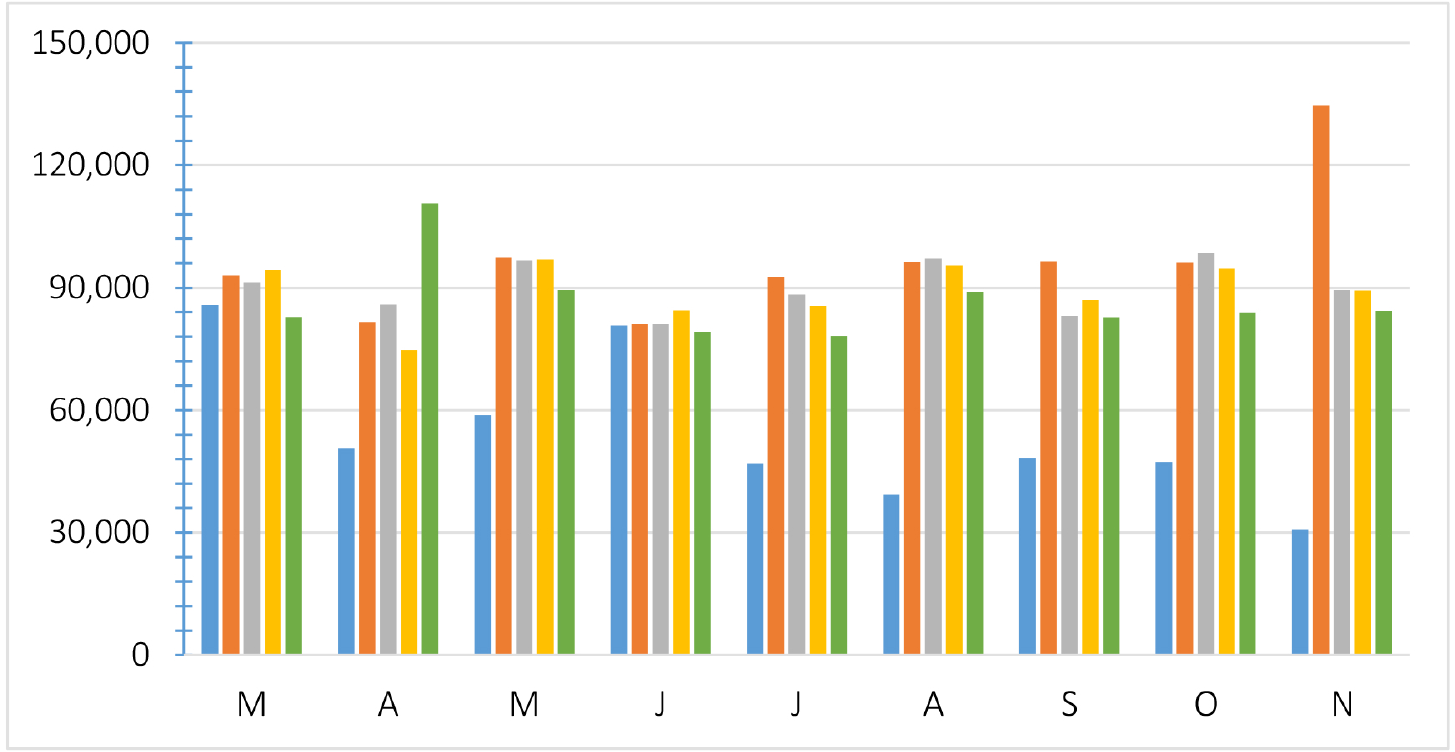}\label{fig:disB}
	}
		\\
	\subfloat[Activity C - absolute frequency by month]{
		\includegraphics[width=0.45\textwidth]{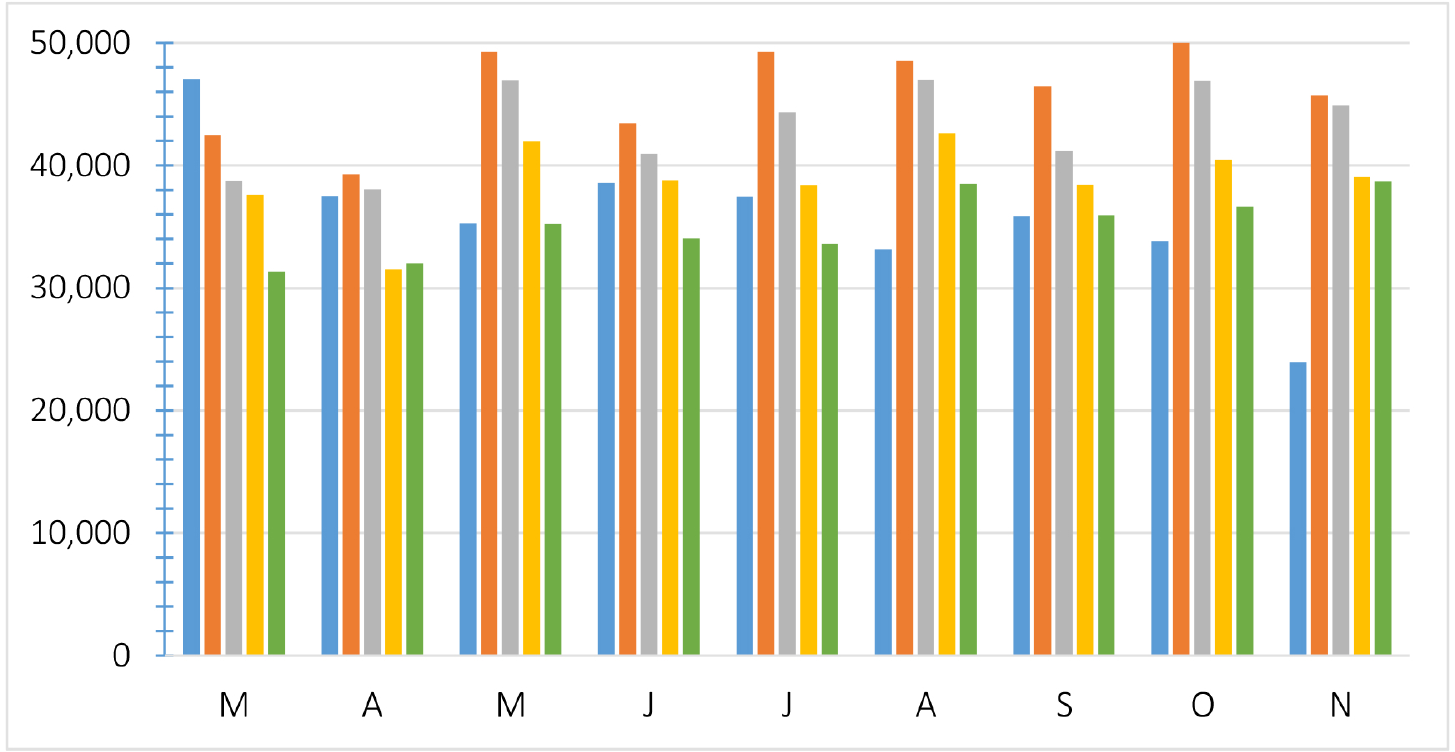}\label{fig:disC}
	}
	\hfill
	\subfloat[Activity D - absolute frequency by month]{
		\includegraphics[width=0.45\textwidth]{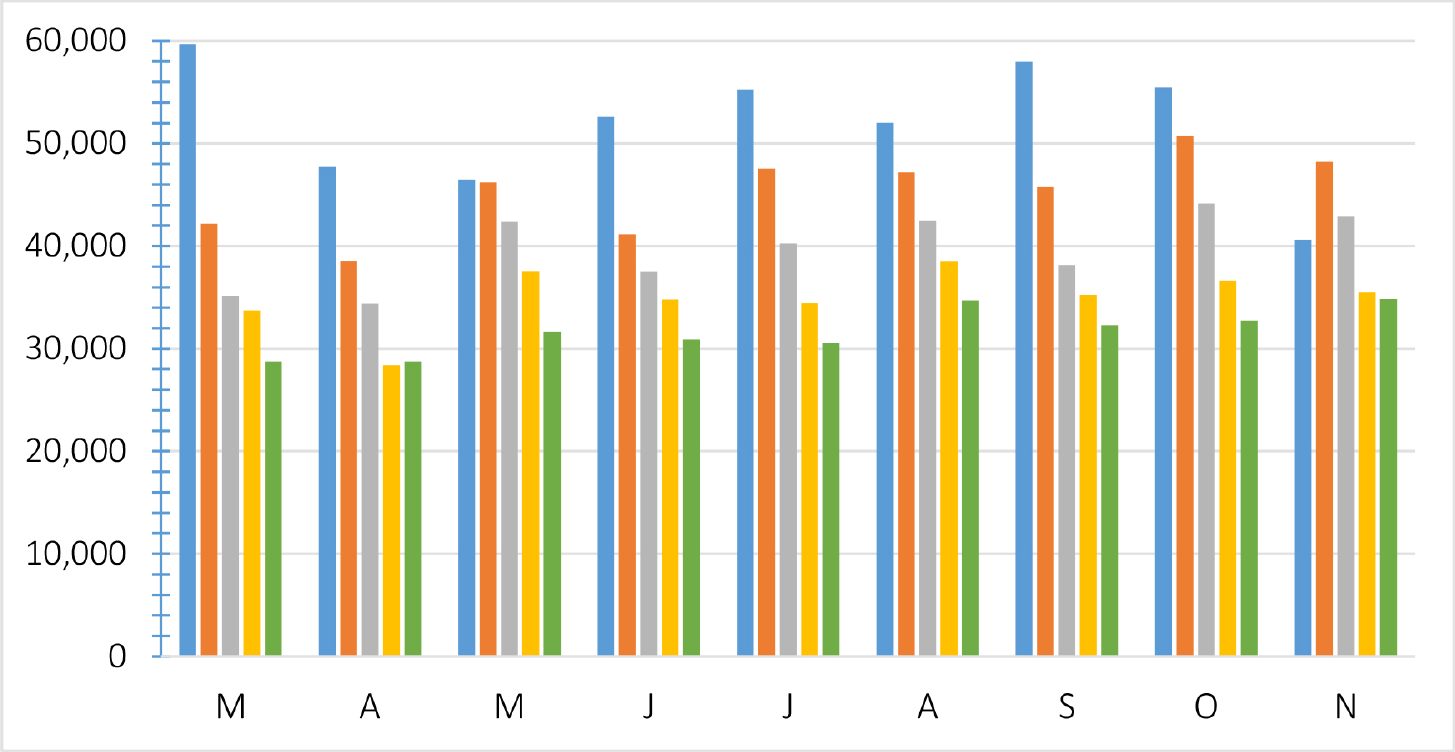}\label{fig:disD}
	}
		\\
	\subfloat[Activity E - absolute frequency by month]{
		\includegraphics[width=0.45\textwidth]{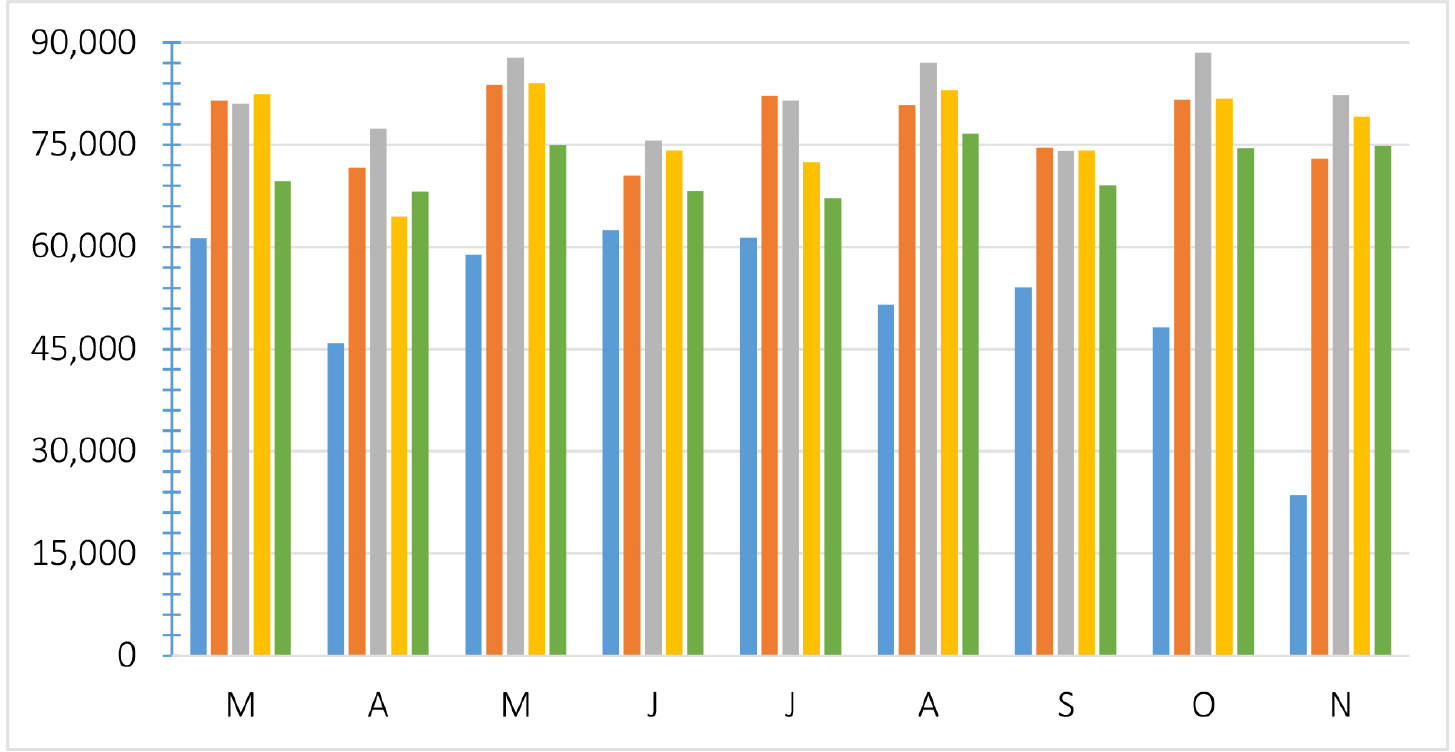}\label{fig:disE}
	}
	\hfill
	\subfloat[Activity F - absolute frequency by month]{
		\includegraphics[width=0.45\textwidth]{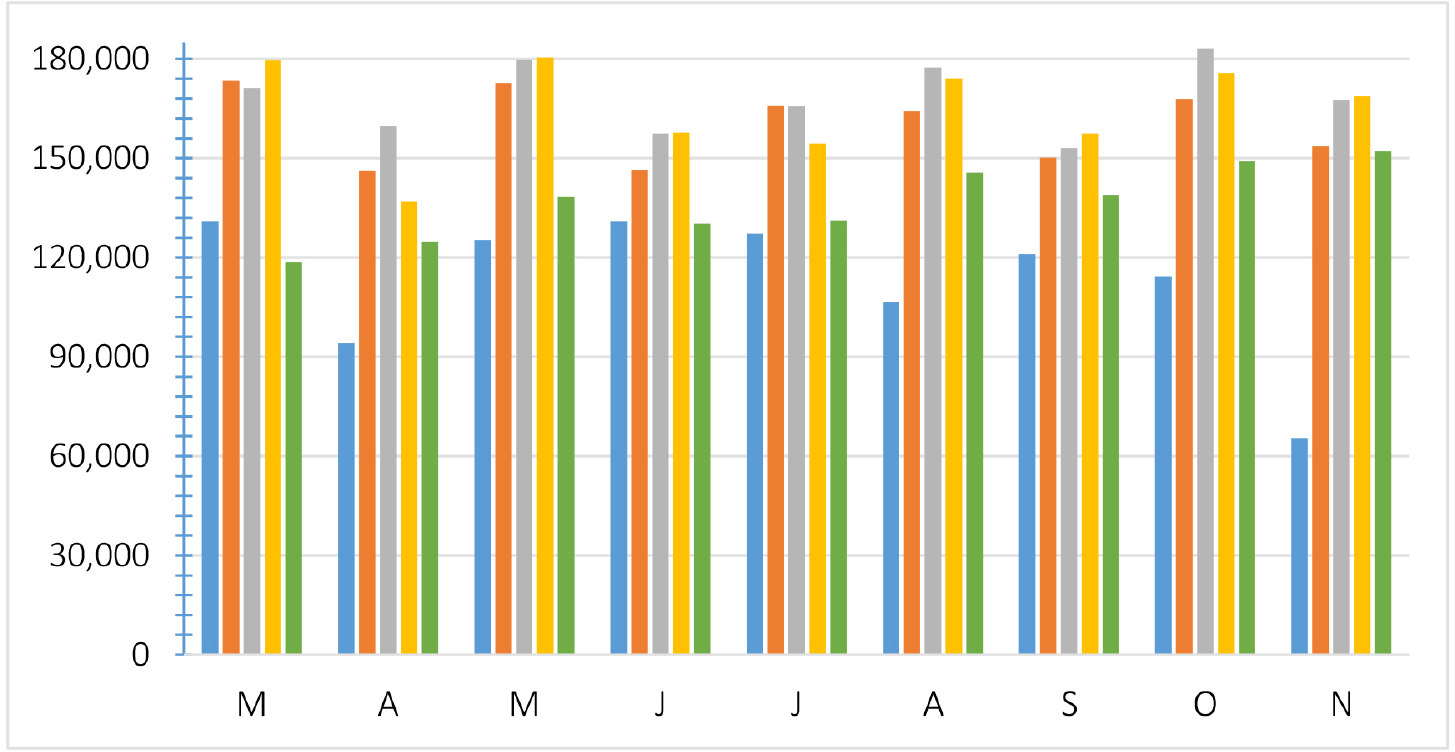}\label{fig:disF}
	}
		\\
	\subfloat[Activity G - absolute frequency by month]{
		\includegraphics[width=0.45\textwidth]{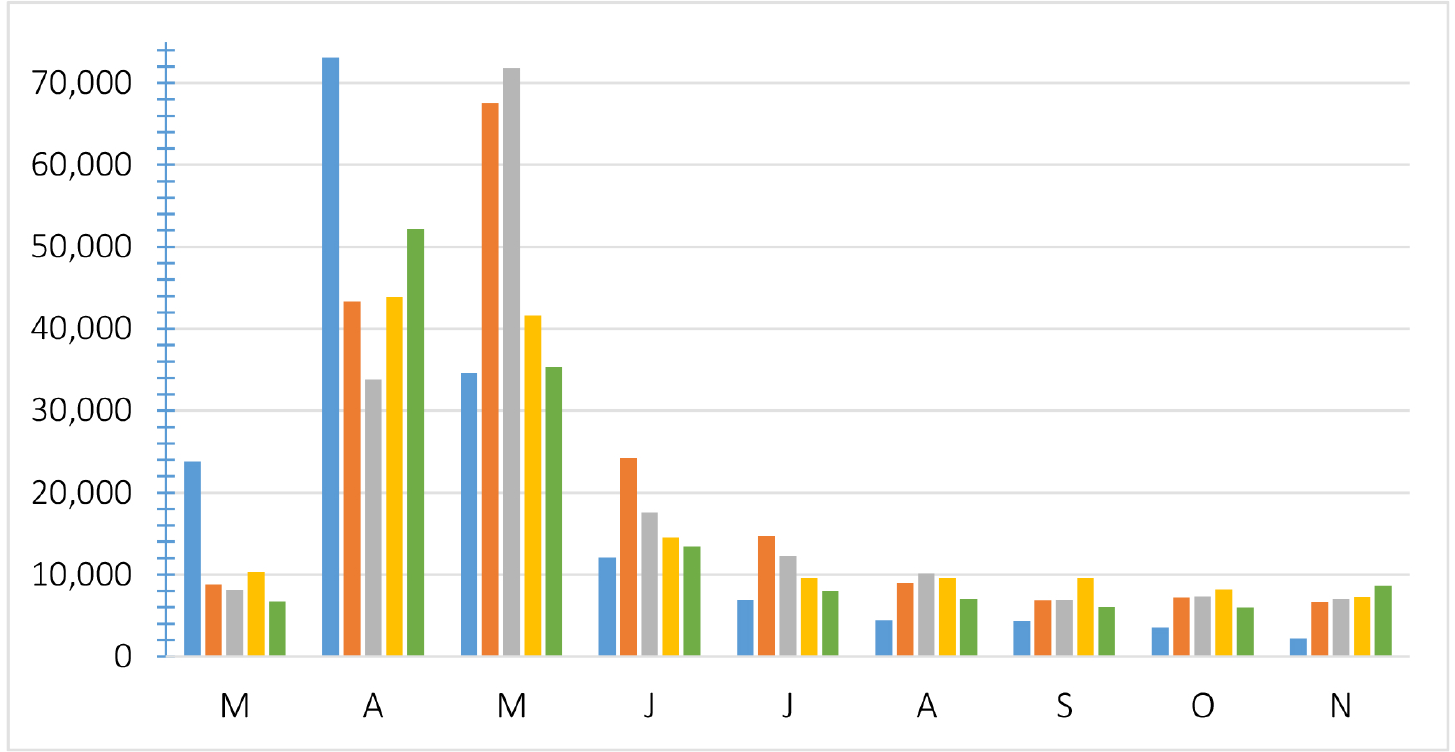}\label{fig:disG}
	}
	\hfill
	\subfloat[Activity absolute frequency changes in 2020]{
		\includegraphics[width=0.45\textwidth]{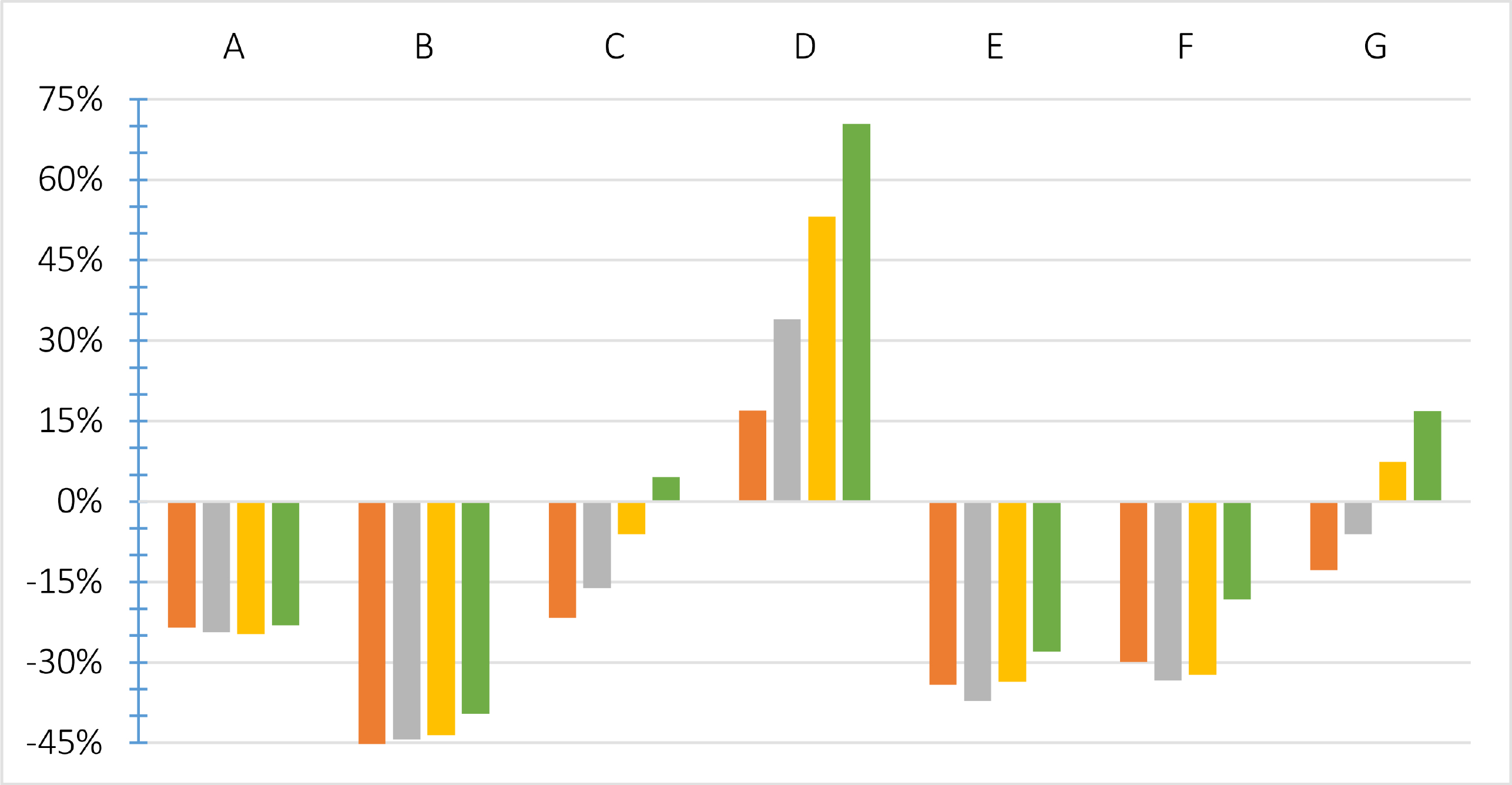}\label{fig:changes}
	}
	\vspace{-4mm}
\caption{Activity frequencies, graphical comparison across years 2020-2016. A = GP Visit, B = GP records measurement, C = medication prescribed, D = medication refill, E = lab/imaging referral, F = test results recorded, G = vaccination administered/recorded.}
\label{fig:plots}
\end{figure}

We explored the distribution of the activities over time, their frequencies, and how they varied over the five years. This information is shown in Figure~\ref{fig:plots}. Figure~\ref{fig:absfreq} and~\ref{fig:relfreq} show the absolute and relative frequencies of each of the seven activities over the five years; 
Figure~\ref{fig:disA} to~\ref{fig:disG} show the absolute frequency of each activity over time, month by month; and Figure~\ref{fig:changes} shows the changes in absolute frequency of each of the activities in 2020, compared to the previous four years. From this data, we can observe the following.

\begin{observation}\label{obs3}
In 2020, the relative frequency of activity $B$ (GP records a measurement) dropped to 9.0\% from an average of 12.1\%. Although this seems a small variation, we note that in 2019, 2018, 2017, and 2016, the relative frequency of activity $B$ was remarkably stable at 12.2\%, 12.0\%, 12.1\%, and 12.2\% (respectively).
\end{observation}

\begin{observation}\label{obs4}
In 2020, the relative frequency of activity $D$ (GP prescribes a refill) increased to 9.2\% from an average of 5.0\%. Also in this case, we note that in 2019, 2018, 2017, and 2016, the relative frequency of activity $D$ was somewhat stable at 5.8\%, 5.1\%, 4.6\%, and 4.4\% (respectively).
\end{observation}

\begin{observation}\label{obs5}
In 2020, the variation in the absolute frequency of activity $G$ (vaccinations are administered/recorded) is remarkably low, in fact, it decreased of only 12.8\% and 6.1\% -- compared to 2019 and 2018, and it increased of 7.4\%	and 16.9\% -- compared to 2017 and 2016. Furthermore, the absolute frequency of activity $G$ is concentrated in the months of March and April, in contrast with the other years, where activity $G$ is mostly observed in April and May.
\end{observation}

Observation~\ref{obs3} can be straightforwardly interpreted. Given that activity $B$ represents a GP taking and recording a measurement of the patient (e.g.\ measuring and recording the patient blood pressure), its decrease can relate to the actual implementation of safety measures -- GP doctors may have avoided to interact with the patients unless strictly necessary.

Observation~\ref{obs4} represents an increase in medication refills. In particular, looking at Figure~\ref{fig:disD}, which captures the activity $D$ distribution over the nine months, we note a clear spike in March, April, June, July, and September. This can relate to an overstocking of drugs by patients that could not risk to run out of their medications. We remind that, during the early COVID-19 pandemic, overstocking was a phenomenon observed across a variety of products from food to toilet paper, known also as \emph{panic buying}~\cite{arafat2020psychological}. However, taking into account the changes of absolute frequency for activity $D$ (see Figure~\ref{fig:changes}), we can observe that drug prescriptions have increased steadily in the past four years with an average increase of 14.2\%. Given that also a similar trend can be observed for activity $C$ (capturing a first-time drug prescription), we cannot conclude that the increase observed in activity $D$ derived exclusively from the COVID-19 pandemic context.

Lastly, Observation~\ref{obs5} is probably the most interesting one, also because it is in contrast with research findings of similar studies conducted in different geographical areas during the equivalent seasons~\cite{santoli2020effects,lassi2021impact}. The data clearly shows that vaccinations were not substantially impacted in 2020, with a decrease in absolute frequency that is lower than the one of other activities (see Figure~\ref{fig:changes}). Leaving aside medication-related activities (i.e., activities C and D), other activities reported an absolute frequency drop of between 23.9\% (on average for activity $A$--GP Visit) and 43.2\% (on average for activity $B$--measurement). While activity $G$ (vaccinations) reported a maximum absolute frequency drop of 12.8\% (compared to 2019) and an average drop of 1.3\%. If we consider this in light of the total drop of the activities observed in 2020 (23.6\% on average, see Table~\ref{tab:logs} -- total events), the drop of vaccination activities is well below the average. In addition, there is a noticeable shift in the vaccination timeline for the 2020 year, bringing  the vaccinations forward of one month. Observation~\ref{obs5} set a direction for additional analysis, which led us to additional findings that we will discuss in depth in Section~\ref{sec:discussion}.

\subsection{Challenge 1 -- Imprecise timestamps}\label{sec:challenge1}

Until now, we have described and analysed the data in general terms. Although we approached it from a process perspective, identifying the process activities and their execution over time, we have not discussed nor analysed the process behaviour, i.e., how such activities follow one another, and what their execution leads to. To analyse the process behaviour, process mining methodologies and tools often rely on directly-follows relations~\cite{ProcessMiningBook} (see Definition~\ref{def:dfr}), especially, for automated discovery of process models~\cite{augusto2021optimization}, and for process variant analysis~\cite{bolt2018process,nguyen2018multi,taymouri2020business}.

Recalling the event log definition (Definition~\ref{def:log}), given that the order of the events in an event log is imposed by the order of their timestamps, incorrect or imprecise timestamps can have a significant (negative) impact on the identification of directly-follows relations and, consequently, on the output of process mining tools that rely on directly-follows relations. This is a well-known problem in the field of process mining~\cite{SuriadiAHW17,bose2013wanna}, especially in healthcare~\cite{mans2012process}, where activities are documented manually. We recall that also in our case the event timestamps had a day-granularity. 

To give an idea of the issue, let us consider a patient visiting a GP doctor (activity $A$), the doctor measures the blood pressure of the patient (activity $B$), and then prescribes a medication for the first time (activity $C$). The activities order is $\langle A, B, C\rangle$. However, they will be recorded in the information systems having all the same timestamp (i.e., the day of the visit), and not necessarily in the order they have been executed. For example, the fact that the patient has visited the doctor may be recorded at the end of a consultation, and the doctor may log activities $B$ and $C$ after they really occurred (inputting them manually on a computer software). As a result, the actual recording may read as follow $\langle C, A, B\rangle$. The more the activities to be recorded, the more are the users involved in their (manual) logging, the greater is the amount of errors.

Past research studies in process mining have addressed the problem of cleaning (or repairing) imprecise timestamps and timestamps errors ~\cite{rogge2013improving,WangSLZP15,song2016cleaning,conforti2020automatic}, however, three of the proposed methods require as input a reference process model~\cite{rogge2013improving,WangSLZP15,song2016cleaning}, while the method of Conforti et al.~\cite{conforti2020automatic} requires to have at least a subset of the events recorded in the event log that are not affected by imprecise timestamps. In our case, we could not rely on any of these existing methods, missing their requirements. 

While recent work~\cite{martin2020recommendations} called for improving the quality of the data captured by healthcare information systems, with the goal to fix the problem at its root, we would like to highlight the opportunity (and the need) for additional research addressing the problem of automated repairing and the cleaning of event log data errors -- especially timestamps.

To continue our analysis and ensure the most reliable outcome, we devised an effective solution to deal with the imprecise timestamps.
We imposed a standard order among the activities (matching the alphabetical order of their labels, see Table~\ref{tab:actmap}), and we reordered the events in the event log based on two attribute values: the event timestamp \emph{and} the event activity. The latter attribute used as a tie-breaker on timestamps equality. 

For example, let us consider the following sequence of events $\langle \event_1, \event_2, \event_3, \event_4, \event_5\rangle$, and let us assume that the five events ($\event_1$ to $\event_5$) have all the same timestamp and that the corresponding sequence of activities is $\langle D, A, G, F, E\rangle$. In such a case, we would reorder the events as $\langle \event_2, \event_1, \event_5, \event_4, \event_3\rangle$, yielding the sequence of activities $\langle A, D, E, F, G\rangle$. Note that the event IDs do not play a role in the ordering. Events having the same ID will be ordered correctly, while events having different IDs would not be affected by the reordering.

Our solution is based on the idea that, in most of the scenarios (and especially in healthcare), certain activities have logical order constraints, e.g., a GP doctor cannot take and record a patient blood pressure (activity $B$) if the patient is not attending a visit (activity $A$). Yet, our solution has limitations, given that not all the activities have a logical order constraint, e.g., a patient may be administered a vaccine (activity $G$) either before or after she is prescribed a medication (activity $C$ or $D$). In fact, there are only three strict logical order constraints in our case, and they are:
$A$ before $B$, $C$ before $D$, and $E$ before $F$. The order we imposed satisfies the three constraints, but also enforces others. We note that, while enforcing additional constraints may distort the factual reality, it homogenise the data allowing for a correct and fair comparison. 

To describe the effects of our solution, let us consider two traces $\langle A, B, G \rangle$ and $\langle A, G, B \rangle$, and let us assume that all the events within each trace have the same timestamp. Comparing the two traces as they are would tell us that they are different, but according to the data they are not (i.e., the timestamps are equal, so any order is valid in principle). Enforcing a standard order over the activities as a tie-breaker on timestamp equality ultimately leads to data standardization and a correct interpretation. 

Our approach for fixing imprecise timestamps due to high-level granularity can be generalized to virtually any other context when the objective of the process analysis is the comparison of process variants, so it should not be considered as an ad-hoc approach for our specific scenario. However, we acknowledge that to define the logical order on the activities, the input of domain experts may be required. In our case, we relied on the experience in general practice medicine of the co-authors Dr Capurro and Dr Manski-Nankervis.

Lastly, we note that the time complexity of our approach is linear on the number of events contained in the event log, making it not only effective but also efficient.


\subsection{Challenge 2 -- Unbounded process instances}\label{sec:challenge2}

Once we solved the problem of imprecise timestamps, we focused on the process behaviour, analysing how the process activities follow one another and what their execution leads to.
However, we note that our healthcare process instances do not perfectly fit the traditional definition  of process~\cite{dumas2013fundamentals} (see Definition~\ref{def:process}), because they miss both a specific start event and a specific end event, making these process instances unbounded.

In our context, a patient may consult their GP doctor to discuss several health issues at once, each of them may lead to different outcomes and some of them may never reach an outcome (e.g., a chronic disease, which requires to be indefinitely monitored), forcing the customer to indefinite follow-ups. At the same time, while following health issues up, new health issues may arise. As one can see, the GP day-to-day healthcare process is conceptually unbounded. In particular, when we look at the activities of a patient within a specific timeframe, the first activity we observe is not necessarily the one that started their GP day-to-day healthcare process, and the only way to determine that with 100\% accuracy would be to have a timeframe at least equal to the patient age -- which is an unrealistic requirement for most of the patients.

Existing process mining techniques for automated process discovery and variant analysis (e.g.,~\cite{augusto2018split,vanden2017fodina,bolt2018process,cecconi2021detection}) are not very effective when dealing with unbounded process instances, because by design they would implicitly (and erroneously, in our context) assume the first event of a trace in the input event log to be the start of the process instance, and the last event of a trace to be the end of the process instance. We can, however, identify the most appropriate start and end events given a process instance. This can be achieved by narrowing down the scope of an unbounded process instance, for example, by focusing on a single GP visit or a single health issue/procedure. To do that we devised an algorithm that leverages domain experts knowledge, once again, the co-authors Dr Capurro and Dr Manski-Nankervis. 

We started from the assumption that a process instance should begin with a visit to the GP doctor (i.e., activity $A$), effectively making activity $A$ the only possible start event of a trace. Any subsequent activity different than activity $A$ (i.e., activities $B$ to $G$) is assumed to be a follow-up of the initial visit to the GP doctor. However, when a second activity $A$ is observed for the same process instance, we have to distinguish two cases:
i) the new activity $A$ is a follow-up of the past activities; ii) the new activity $A$ is not related to the past activities (i.e., this would trigger a new process instance). We distinguished the two cases on a time basis. Precisely, if the new activity $A$ is more than six months away from the first observed activity $A$ and more than one month away from the last observed activity of the current process instance, we are in case ii); otherwise, we are in case i). These time thresholds were set empirically following the domain experts.

Algorithm~\ref{alg:gentraces} describes a generalisation of our approach to generate traces from a given event log containing unbounded process instances.
The algorithm takes in input the log ($\elog$), a set of allowed start activities ($\fact$) -- in our case containing only activity $A$, and two time thresholds $\dtz$ and $\dtn$ -- in our case six- and one-month respectively.
Three data structures are initialised (see lines~\ref{alg:map} to~\ref{alg:timen}):
i) a map linking an event ID to its trace ($\traces$) -- representing the collection of traces to output;
ii) a map linking an event ID to the timestamp of the first event in the corresponding trace ($\timestamps_0$);
and iii) a map linking an event ID to the timestamp of the last observed event in the corresponding trace ($\timestamps_n$).
Then, we read the log ($\elog$) one event at a time, starting from its first event ($\event$, line~\ref{alg:mloop}).

If the event ID ($\eventid$) is not yet in the map $\traces$ \emph{and} the event activity ($\eventa$) is in the set $\fact$, we create a new empty trace ($\trace$), we append $\event$ to $\trace$, we add the event ID and the trace to the map $\traces$, we save the timestamp of $\event$ in $\timestamps_0$ and in $\timestamps_n$ (lines~\ref{alg:newtrace1} to~\ref{alg:newtrace2}). 

If $\eventid$ is already mapped in $\traces$ \emph{and} $\eventa$ is not an allowed start activity (line~\ref{alg:case0}), we retrieve the trace linked to the event ID ($\traces(\eventid)$) and we append $\event$ to that trace (line~\ref{alg:case0_append}). Then, we update the timestamp information by overwriting the last observed event timestamp in the map $\timestamps_n$ (line~\ref{alg:case0_overwrite}).

If $\eventid$ is already mapped in $\traces$ \emph{and} $\eventa$ is an allowed start activity (line~\ref{alg:case0}), we distinguish the two possible cases mentioned above. Case ii), if $\eventime$ is less than or equal to $\dtz$ \emph{or} less than or equal to $\dtn$, then we append $e$ to the already existing trace $\traces(\eventid)$ (as just described above -- see lines~\ref{alg:case2} to~\ref{alg:case2e}). Otherwise, Case i), we create a new event ID (that is not present in the event log),\footnote{This can be achieved by manipulating the current $\eventid$.} we link the new event ID to the existing trace in $\traces$ that is mapped to $\eventid$, we create a new empty trace ($\trace$), we append $\event$ to $\trace$, we add the event ID and the trace to the map $\traces$, we save the timestamp of $\event$ in $\timestamps_0$ and in $\timestamps_n$ (lines~\ref{alg:case1_newtrace1} to~\ref{alg:case1_newtrace2}).

Once all the events in the event log have been read, Algorithm~\ref{alg:gentraces} returns the map of event IDs and the corresponding traces. 

Assuming that accessing the maps is a constant-time operation, as it is the case in modern object-oriented programming languages, Algorithm~\ref{alg:gentraces} has a linear time complexity on the number of events contained on the event log.

We note that the information shown in Table~\ref{tab:logs} is the one obtained after the execution of Algorithm~\ref{alg:gentraces}. The column \emph{filtered events} reports the number of events that were removed by applying Algorithm~\ref{alg:gentraces}, i.e., events that are not preceded by an activity $A$. On average, we removed 3.6\% of events from the data, which is a negligible amount.

\begin{algorithm}[hbtp]
{\scriptsize{
	\Input{Event Log $\elog$, Set $\fact$, Integer $\dtz$ , Integer $\dtn$}
	\Output{Traces $\traces$}
    \BlankLine
    \textbf{Map} $\traces \leftarrow \varnothing$\;\label{alg:map}
    \textbf{Map} $\timestamps_0 \leftarrow \varnothing$\;
    \textbf{Map} $\timestamps_n \leftarrow \varnothing$\;\label{alg:timen}
    \BlankLine
    \For{ $\event \in \elog$ }{\label{alg:mloop}
        \eIf{ $(\eventid \notin \traces)$ \emph{AND} $(\eventa \in \fact)$}{\label{alg:newtrace1}
	        \textbf{Create} $\trace$\;
	        \textbf{Append} $\event$ \textbf{to} $\trace$\;
	        \textbf{Add} ($\eventid$, $\trace$) \textbf{to} $\traces$\;
	        \textbf{Add} ($\eventid$, $\eventime$) \textbf{to} $\timestamps_0$\;
	        \textbf{Add} ($\eventid$, $\eventime$) \textbf{to} $\timestamps_n$\;\label{alg:newtrace2}
	    }{ 
            \If{$(\eventid \in \traces)$ \emph{AND} $(\eventa \notin \fact)$}{\label{alg:case0}
	            \textbf{Append} $\event$ \textbf{to} $\traces(\eventid)$\;\label{alg:case0_append}
	            \textbf{Add} ($\eventid$,$\eventime$) \textbf{to} $\timestamps_n$\;\label{alg:case0_overwrite}
            }
            
	        \If{ $(\eventid \in \traces)$ \emph{AND} $(\eventa \in \fact)$}{
	           \eIf{ $(\eventime - \timestamps_0(\eventid) \leq \dtz)$ \emph{OR} $(\eventime - \timestamps_n(\eventid) \leq \dtn)$ }{\label{alg:case2}
	                \textbf{Append} $\event$ \textbf{to} $\traces(\eventid)$\;
	                \textbf{Add} ($\eventid$, $\eventime$) \textbf{to} $\timestamps_n$\;\label{alg:case2e}
	           }{
	                \textbf{ID} $\leftarrow$ generateID($\eventid$)\;\label{alg:case1}
	                \textbf{Add} (ID, $\traces(\eventid)$) \textbf{to} $\traces$\;\label{alg:case1_newtrace1}
	                \textbf{Create} $\trace$\;
	                \textbf{Append} $\event$ \textbf{to} $\trace$\;
	                \textbf{Add} ($\eventid$, $\trace$) \textbf{to} $\traces$\;
	                \textbf{Add} ($\eventid$, $\eventime$) \textbf{to} $\timestamps_0$\;
	                \textbf{Add} ($\eventid$, $\eventime$) \textbf{to} $\timestamps_n$\;\label{alg:case1_newtrace2}
	           }
	        }
	    }
	}
	\BlankLine

    \textbf{return} $\traces$\;
	}}
	\caption{Generate traces from the event log}\label{alg:gentraces}
\end{algorithm}

\subsection{Challenge 3 -- Any process behaviour is allowed}\label{sec:challenge3}

At this stage, we can finally turn our attention to the process behaviour analysis,
by leveraging process mining techniques~\cite{van2015pm}.
Since we are interested in identifying process behavioural differences over five different timeframes (each captured in an event log), the appropriate process mining techniques are in the class of automated process discovery~\cite{augusto2018automated} and process variant analysis~\cite{taymouri2021business}.
Automated process discovery techniques receive in input an event log and automatically produce a process model, which is a graphical representation of the process behaviour, such as a workflow chart, a Petri net, or a BPMN model~\footnote{\url{https://www.bpmn.org/}}. By looking at different process models, it is possible to detect behavioural differences. On the other hand, process variant analysis techniques receive two event logs and automatically produce an artifact that highlights the process behavioural differences. Differences captured by variant analysis techniques are either at control-flow level (i.e., process behavioural differences in terms of executed activities) or at performance level (i.e., differences in the execution/hand-over times of/between the process activities). 
From both classes of techniques, we selected three state-of-the-art tools, based on previous studies evaluations~\cite{augusto2018automated,taymouri2020business,cecconi2021detection} which are:
Fodina~\cite{vanden2017fodina}, Inductive Miner~\cite{leemans2014infrequent}, Split Miner~\cite{augusto2018split}, and their metaheuristics optimization variants~\cite{augusto2021optimization} (for automated process discovery); and process comparator~\cite{bolt2018process}, fingerprints-based variant analysis~\cite{taymouri2020business}, and variant analysis via declarative rules~\cite{cecconi2021detection} (for variant analysis).

We attempted to discover a process model from each of the five event logs, by running each of the three automated process discovery tools. The models we obtained were not structurally complex (spaghetti-models), but they showed that any behaviour was allowed -- with minimal constraints and many cyclical patterns. This finding highlights that the GP day-to-day healthcare processes have a behavioural degree of freedom that is not comparable with most business processes, allowing a vast amount of different behaviour to be executed and repeated over time. 

\begin{figure}[H]
	\centering
	\subfloat[Inductive Miner~\cite{leemans2014infrequent} process model (GP19 and GP20 event logs)]{
		\includegraphics[width=0.35\textwidth]{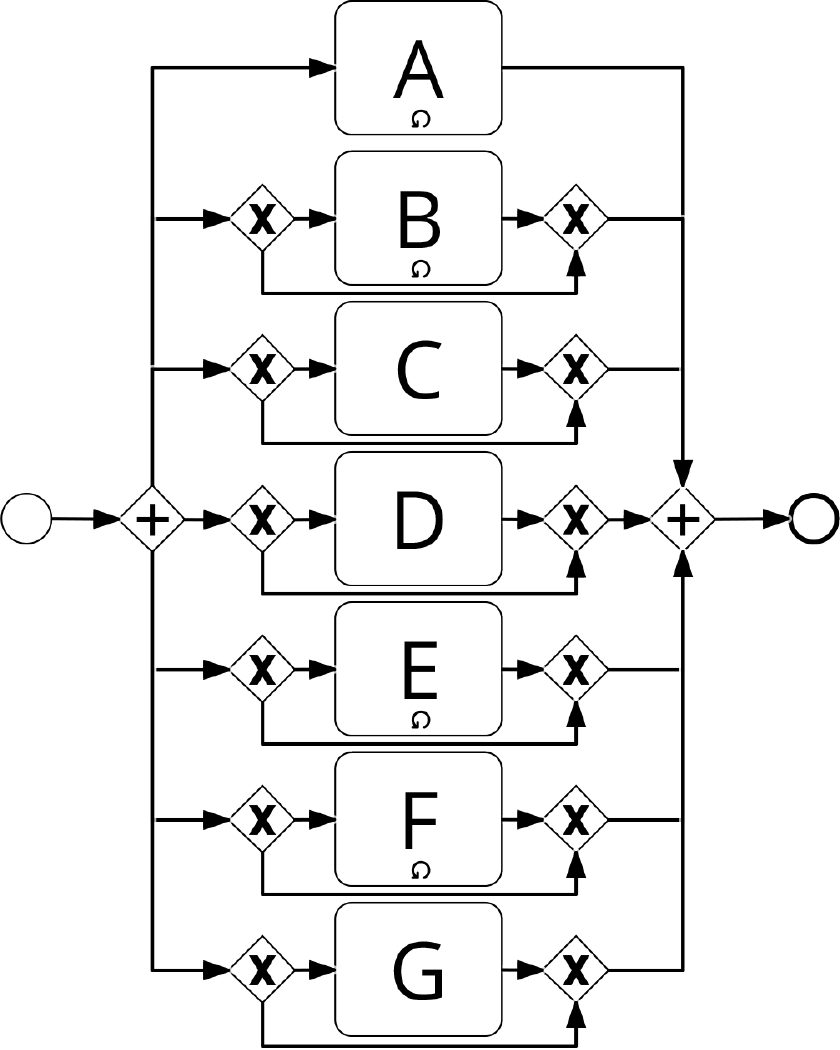}\label{fig:im201920}
	}
	\hfill
	\subfloat[Split Miner~\cite{augusto2018split} process model (GP19 event log)]{
		\includegraphics[width=0.58\textwidth]{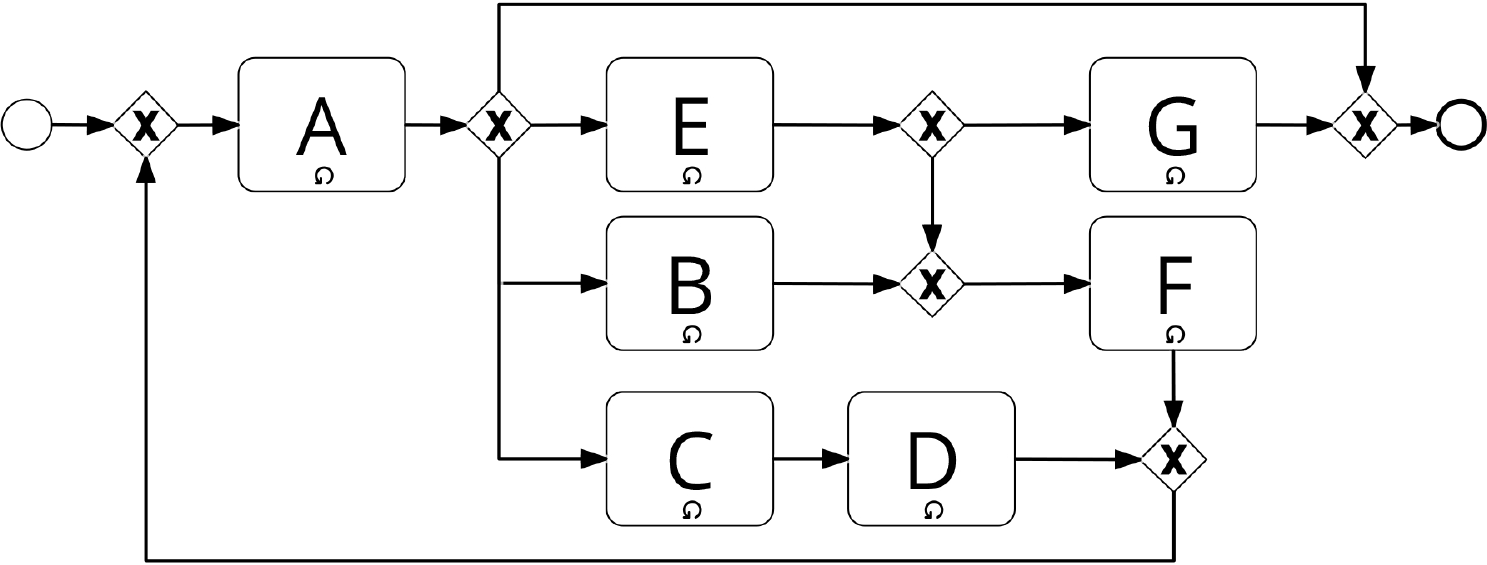}\label{fig:sm2019}
	}
		\\
	\subfloat[Split Miner~\cite{augusto2018split} process model (GP20 event log)]{
		\includegraphics[width=0.58\textwidth]{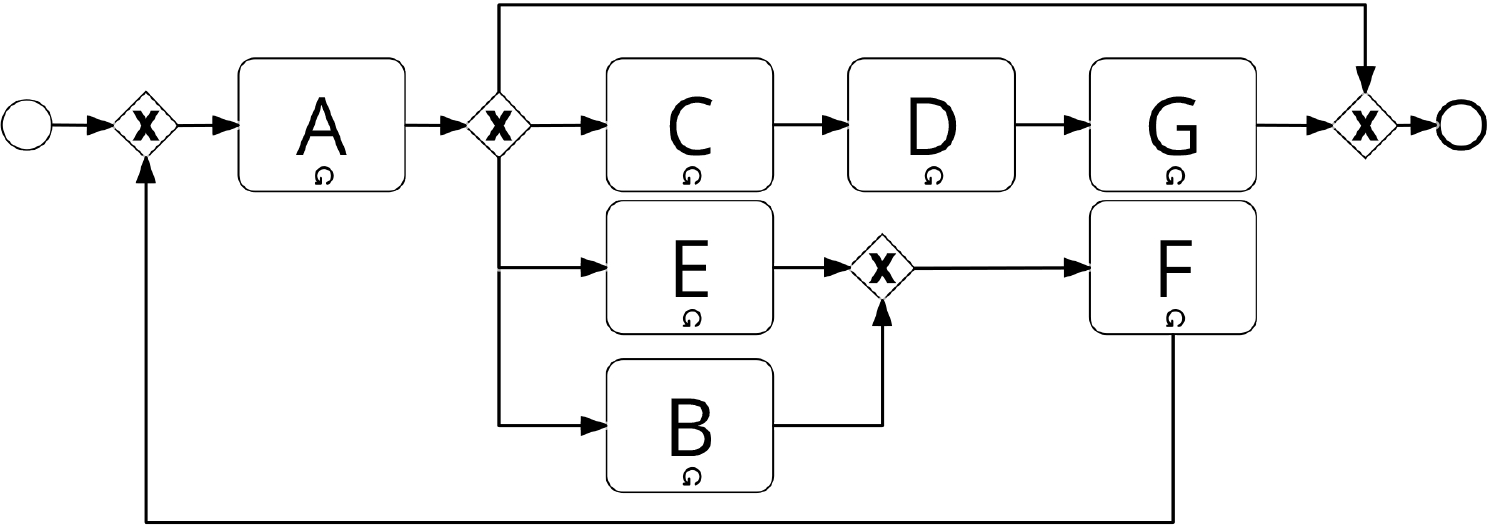}\label{fig:sm2020}
	}
	\vspace{-4mm}
\caption{Automatically discovered process models~\cite{augusto2018split,leemans2014infrequent}, from the GP20 and GP19 event logs}
\label{fig:procmodels}
\end{figure}

As an example, Figure~\ref{fig:im201920} and~\ref{fig:sm2020} show the models discovered by Inductive Miner and Split Miner from the GP20 log, while Figure~\ref{fig:im201920} and~\ref{fig:sm2019} show the models discovered by the same techniques but from the GP19 log. The process models discovered by Split Miner are almost identical (with a small variation involving activity $G$), and they allow for much repetitive and variable behaviour over the set of seven activities. While those discovered by Inductive Miner are in fact identical (we discovered the same model from the two event logs), and they allow for an even wider range of behaviour.~\footnote{For space reason, we do not show all the process models we discovered, since they are similar to the ones we commented.}

Successively, we ran the variant analysis. In this case, we experienced some technical limitations of the existing tools, which future research should consider addressing.
In particular, the tool of Bolt et al.~\cite{bolt2018process} was not able to process the input data -- yielding exceptions.\footnote{We contacted the authors two times, but they were not able to provide support in addressing the technical issues.} The tool of Taymouri et al.~\cite{taymouri2020business} either was unable to provide an output within a two hours timeout~\footnote{Which was a hard timeout imposed by the secure environment we were operating.}, or it was not able to identify statistically significant differences. Lastly, the tool of Cecconi et al.~\cite{cecconi2021detection} was the only one that returned a valid output within the timeout, however, it that was possible only by applying its embedded filtering algorithm -- which allowed us to focus only on the most frequent process behaviour. The top-10 differences identified are reported in Table~\ref{tab:dec-compare}, for instance, the output of the process variant analysis of the GP20 and GP19 logs support and refines Observation~\ref{obs3} and Observation~\ref{obs4} (see Section~\ref{sec:dataset}), highlighting a decrease in the number of observations of activity $B$ in the healthcare processes of 2020 (see Table~\ref{tab:dec-compare}, rows 5, 7 and 10), and an increase in the number of observations of activity $D$ in the healthcare processes of 2020 (see Table~\ref{tab:dec-compare}, rows 1-4 and row 9). Similar results were obtained when comparing the data from the 2020 against the data from the 2018, 2017, and 2016.

\begin{table}[htbp]
  \centering
  {\scriptsize{
  \caption{ Top-10 differences between healthcare processes in 2020 (\emph{Variant X}) and 2019 (\emph{Variant Y})~\cite{cecconi2021detection}	}\label{tab:dec-compare}
  \vspace{-4mm}
\begin{tabular}{l|l}
    
    \hline
1	&	In Variant X	it is	16.70\%	more likely than	Variant Y	that if	[B]	occurs, also	[D]	occurs\\\hline
2	&	In Variant X	it is	16.59\%	more likely than	Variant Y	that if	[E]	occurs, also	[D]	occurs\\\hline
3	&	In Variant X	it is	16.32\%	more likely than	Variant Y	that if	[F]	occurs, also	[D]	occurs\\\hline
4	&	In Variant X	it is	11.90\%	more likely than	Variant Y	that if	[A]	occurs, also	[D]	occurs\\\hline
5	&	In Variant Y	it is	11.36\%	more likely than	Variant X	that if	[F]	occurs, also	[B]	occurs\\\hline
6	&	In Variant Y	it is	11.29\%	more likely than	Variant X	that if	[D]	occurs, also	[C]	occurs\\\hline
7	&	In Variant Y	it is	11.11\%	more likely than	Variant X	that if	[E]	occurs, also	[B]	occurs\\\hline
8	&	In Variant X	it is	10.73\%	more likely than	Variant Y	that	[D]	occurs in a process instance\\\hline
9	&	In Variant X	it is	10.71\%	more likely than	Variant Y	that if	[C]	occurs, also	[D]	occurs\\\hline
10	&	In Variant Y	it is	10.47\%	more likely than	Variant X	that if	[A]	occurs, also	[B]	occurs\\\hline
    
\end{tabular}

    }}
\end{table}

Given that the selected process mining techniques struggled to deal with the variety of behaviour captured in the logs under analysis, we took a step back, and decided to review the behaviour recorded in each of the five event logs by visualising their \emph{directly-follows graphs}.

For reasons of space, clarity, and simplicity, we report the DFG of only two event logs (GP20 and GP19) and in their matrix form, where each matrix row (and column) represents a node of the DFG -- i.e., an activity; and each cell of the matrix captures the frequency of the edge between the two nodes -- i.e., how many times we observe in the event log a directly-follows relation between two activities. 

\begin{table}[htbp]
  \centering
  {\scriptsize{
  \caption{Directly-follows graph in matrix form - 2020 process}\label{tab:dfg2020}%
  \vspace{-4mm}
    \begin{tabular}{r|rrrrrrr}
            \hline
            \textbf{Activity}
          & \textbf{A} 
          & \textbf{B} 
          & \textbf{C} 
          & \textbf{D} 
          & \textbf{E} 
          & \textbf{F} 
          & \textbf{G} \\\hline
          
    \textbf{A} & 916053 & 202147 & 263395 & 168861 & 179713 & 137296 & 111189 \\
    \textbf{B} & 108522 & 60445 & 33589 & 16939 & 20640 & 157124 & 10187 \\
    \textbf{C} & 29990 & 3903  & 64    & 257958 & 9254  & 3929  & 2196 \\
    \textbf{D} & 217229 & 29634 & 406   & 972   & 76571 & 30394 & 16982 \\
    \textbf{E} & 126835 & 40963 & 2343  & 1589  & 8439  & 222635 & 11616 \\
    \textbf{F} & 298881 & 87045 & 9573  & 4856  & 124704 & 376932 & 2143 \\
    \textbf{G} & 89601 & 18193 & 102   & 152   & 3750  & 12611 & 99 \\\hline
    \end{tabular}%
    }}
  
\end{table}%

\begin{table}[htbp]
  \centering
  {\scriptsize{
  \caption{Directly-follows graph in matrix form - 2019 process}\label{tab:dfg2019}%
  \vspace{-4mm}
    \begin{tabular}{r|rrrrrrr}
            \hline
            \textbf{Activity}
          & \textbf{A} 
          & \textbf{B} 
          & \textbf{C} 
          & \textbf{D} 
          & \textbf{E} 
          & \textbf{F} 
          & \textbf{G} \\\hline
          
    \textbf{A} & 1190430 & 449745 & 299143 & 70762 & 337449 & 122043 & 121266 \\
    \textbf{B} & 219968 & 107579 & 78582 & 13069 & 74732 & 221429 & 12921 \\
    \textbf{C} & 46754 & 11415 & 39    & 294829 & 26567 & 3681  & 4079 \\
    \textbf{D} & 179501 & 25863 & 145   & 456   & 88901 & 11993 & 15318 \\
    \textbf{E} & 109527 & 72731 & 1078  & 519   & 19149 & 412086 & 17438 \\
    \textbf{F} & 485141 & 120865 & 16046 & 6182  & 88711 & 556912 & 5921 \\
    \textbf{G} & 108491 & 19419 & 77    & 101   & 7483  & 15293 & 185 \\\hline
    \end{tabular}%
    }}
\end{table}%

Tables~\ref{tab:dfg2020} and~\ref{tab:dfg2019} report the DFGs in matrix form of the event logs GP20 and GP19 (respectively). We note that any directly-follows relation can be observed in two DFGs. Although some of them are rare (e.g., $C \rightarrow C$, with a frequency in the order of hundreds), the vast majority can be observed with a frequency in the order of thousands. The DFGs of the event logs (GP18, GP17, and GP16) are very similar to the two we reported here, and this clearly highlights that any behaviour is allowed in the process under analysis, across the five event logs.

The extent of behavioural variability we are observing is a major cause of strain for state-of-the-art process mining techniques, which can disarm them. An alternative would be to remove some behaviour by applying filtering techniques~\cite{ConfortiRH17,DBLP:journals/jiis/TaxSA19,DBLP:conf/bpm/SaniZA17}, however, we recall that all the automated process discovery algorithms that we used already apply a filter~\cite{vanden2017fodina,leemans2014infrequent,augusto2018split}, as well as two of the three variant analysis approaches~\cite{bolt2018process,cecconi2021detection}. 

While in a business context some infrequent behaviour may be a violation of compliance or internal business rules, in our context, all behaviour is actually allowed. As such, we are not interested in removing behaviour, but rather change our focus, and consider only a portion of behaviour that can be fruitfully analysed.

With that in mind, we considered only the most frequent behaviour. Table~\ref{tab:topbeh} reports the top-20 most frequent traces that we could observe in each of the five event logs. Scanning carefully through Table~\ref{tab:topbeh}, we notice that in 2020, traces containing the activity $G$ were more frequent than other years (for clarity, we reported these traces in Table~\ref{tab:topbehG}).
In 2020, not only the traces containing the activity $G$ were more frequent, but they accounted for the 25\% percent of the most frequent behaviour (5 traces out of 20). This finding is remarkable, and when paired it with Observation~\ref{obs5} (discussed in Section~\ref{sec:observations}) clearly hints to a variation in the behaviour involving vaccinations during the 2020.
A similar reasoning can also be done for traces containing the activity $D$ (see $\langle A, D\rangle$, across the five logs). Further investigation of the most frequent process behaviour may reveal several additional differences, but within the scope of this study, we decided to investigate the specific behavioural difference related to activity $G$ and its traces among the top-20.

Our process mining analysis highlights two limitations of the state-of-the-art process mining techniques that we used in this study:
\begin{itemize}
    \item[1.] Process mining techniques for automated process discovery and process variant analysis suffer of scalability and/or quality issues when they deal with too much and too variable behaviour.
    
    \item[2.] Automated process discovery techniques try to capture as much behaviour as possible from the event log, filtering infrequent behaviour only when it is strictly necessary to either simplify the process model or increase its accuracy. However, depending on the context, one may be interested in capturing very little process behaviour from the event log -- requiring special filtering techniques.
\end{itemize}

While limitation 1 is ground for future research directions and studies. Limitation 2, at the moment, can be addressed manually, by applying ad-hoc filters of the process behaviour recorded in the event logs (as we did). The best ad-hoc filters must be identified by domain experts, often on a trial-and-error basis, and applied either via ProM plugins or commercial tools such as Celonis, Disco, or Apromore (which we used). Future process mining techniques should allow the user to automatically design such filters, without relying on domain experts knowledge. For instance, by automatically analysing the outputs of a set of process mining techniques (e.g., both automated process discovery and variant analysis) -- as we did manually.
\begin{figure}[H]
	\centering
	\subfloat[Year 2016, 2017, and 2018]{
		\includegraphics[width=0.40\textwidth]{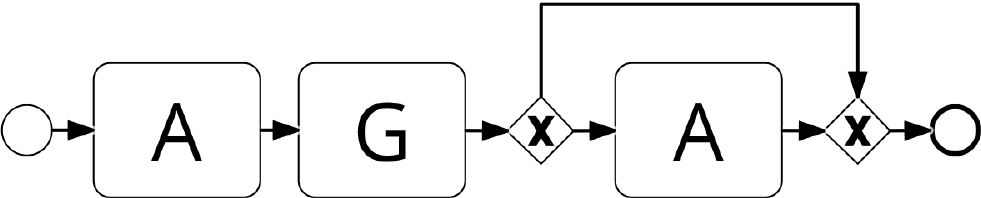}\label{fig:vacproc201618}
	}
		\hfill
	\subfloat[Year 2019]{
		\includegraphics[width=0.55\textwidth]{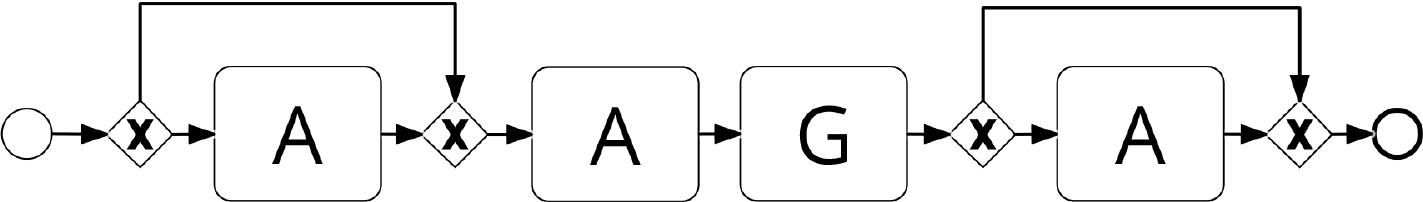}\label{fig:vacproc2019}
	}
		\\
	\subfloat[Year 2020]{
		\includegraphics[width=0.90\textwidth]{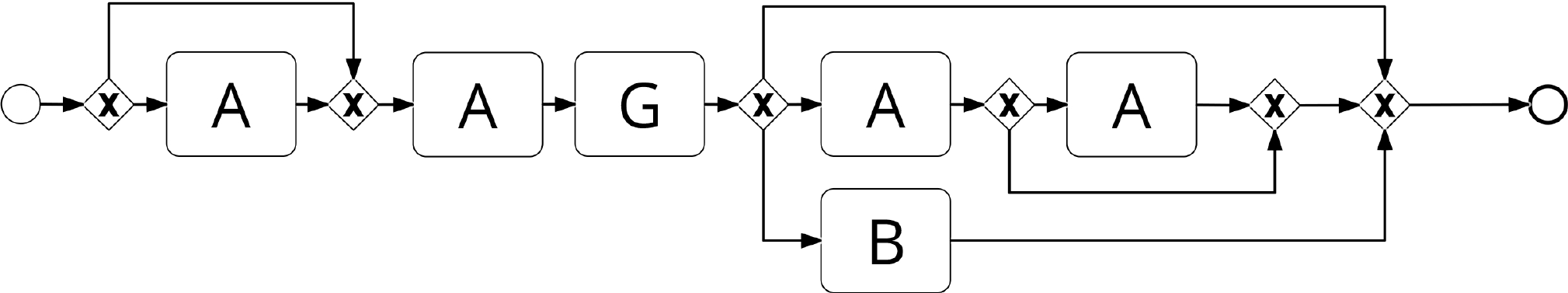}\label{fig:vacproc2020}
	}
	\vspace{-4mm}
\caption{Vaccination process models (most frequent behaviour only)}
\label{fig:vac-processes}
\end{figure}

Once we narrowed our attention down to only the most frequent traces containing a vaccination event (activity $G$), we could easily discover a clear and simple process model for each of the five years and identify the differences, Figures~\ref{fig:vacproc201618} to~\ref{fig:vacproc2020} show the process models. We note that in 2016, 2017, and 2018, the most frequent vaccination process is the same (Figure~\ref{fig:vacproc201618}), changes in this process are minimal in 2019 (Figure~\ref{fig:vacproc2019}), and substantial in 2020 (Figure~\ref{fig:vacproc2020}). 

\begin{figure}[H]
	\centering
	\subfloat[Activity(1) $A$ - Optional]{
		\includegraphics[width=0.3\textwidth]{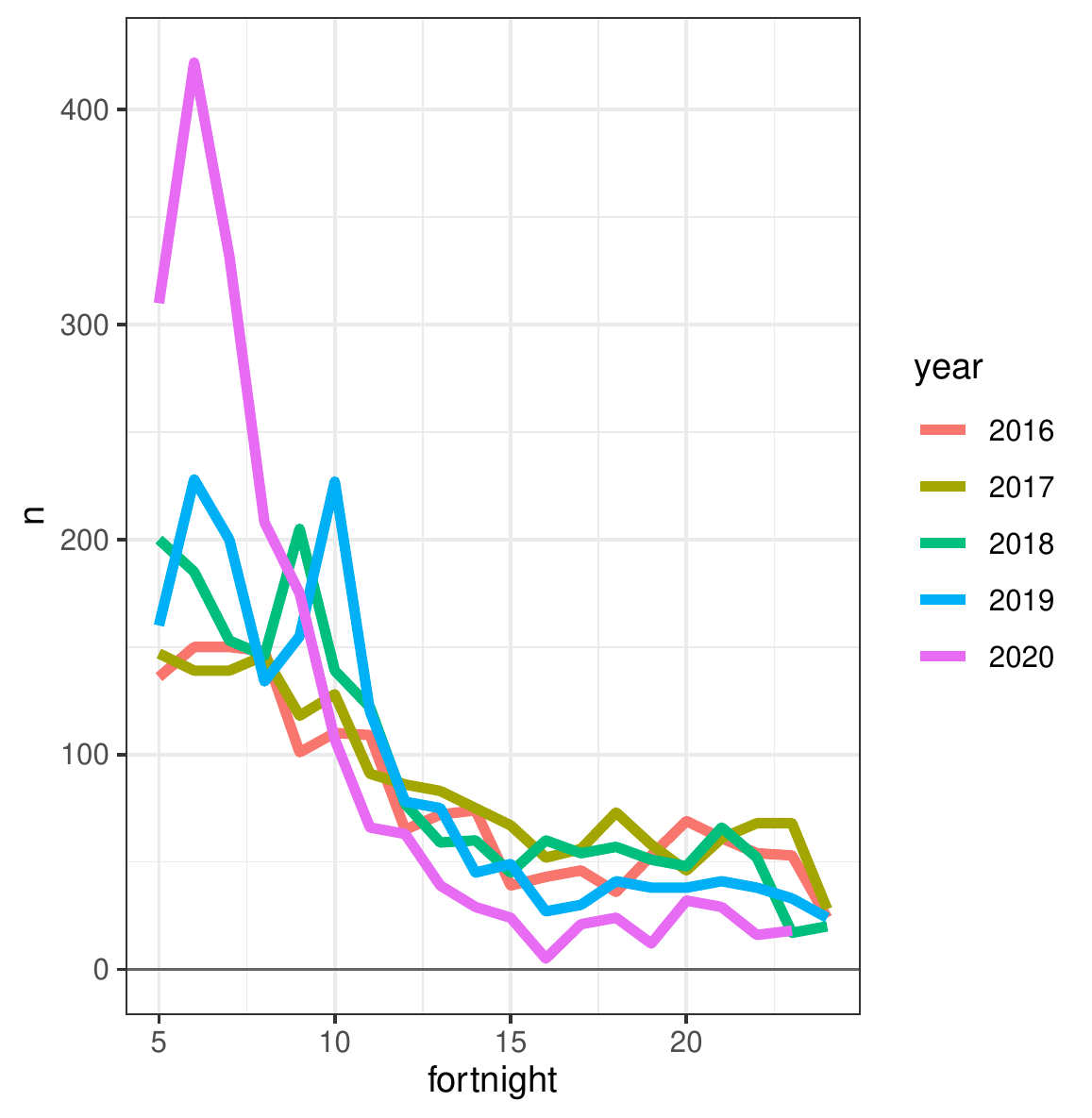}\label{fig:actA1}
	}
		\hfill
	\subfloat[Activity(2) $A$ - Compulsory]{
		\includegraphics[width=0.3\textwidth]{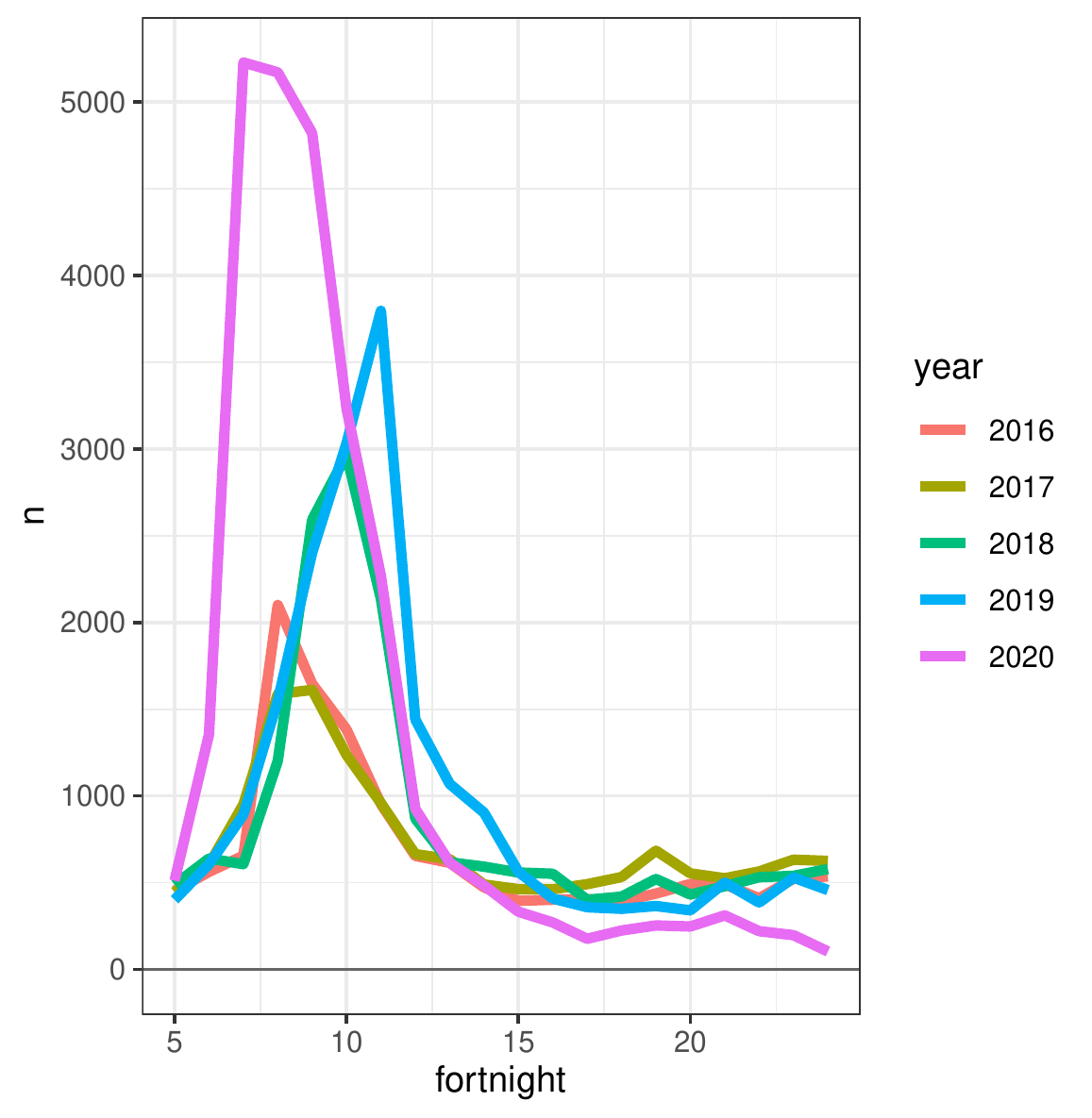}\label{fig:actA2}
	}
		\hfill
	\subfloat[Activity(3) $G$ - Compulsory]{
		\includegraphics[width=0.3\textwidth]{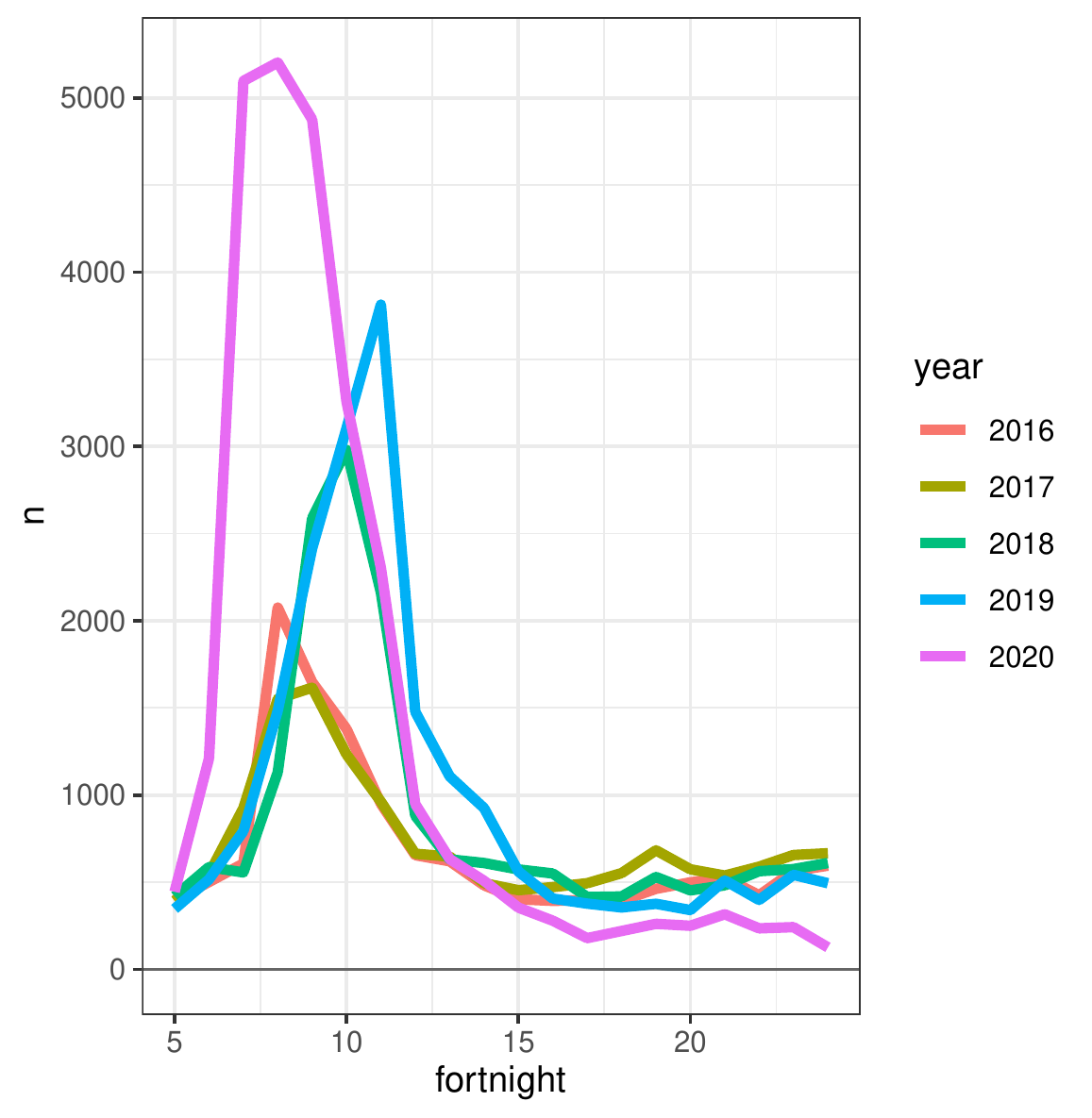}\label{fig:actG3}
	}
		\\
	\subfloat[Activity(4) $A$ - Optional]{
		\includegraphics[width=0.3\textwidth]{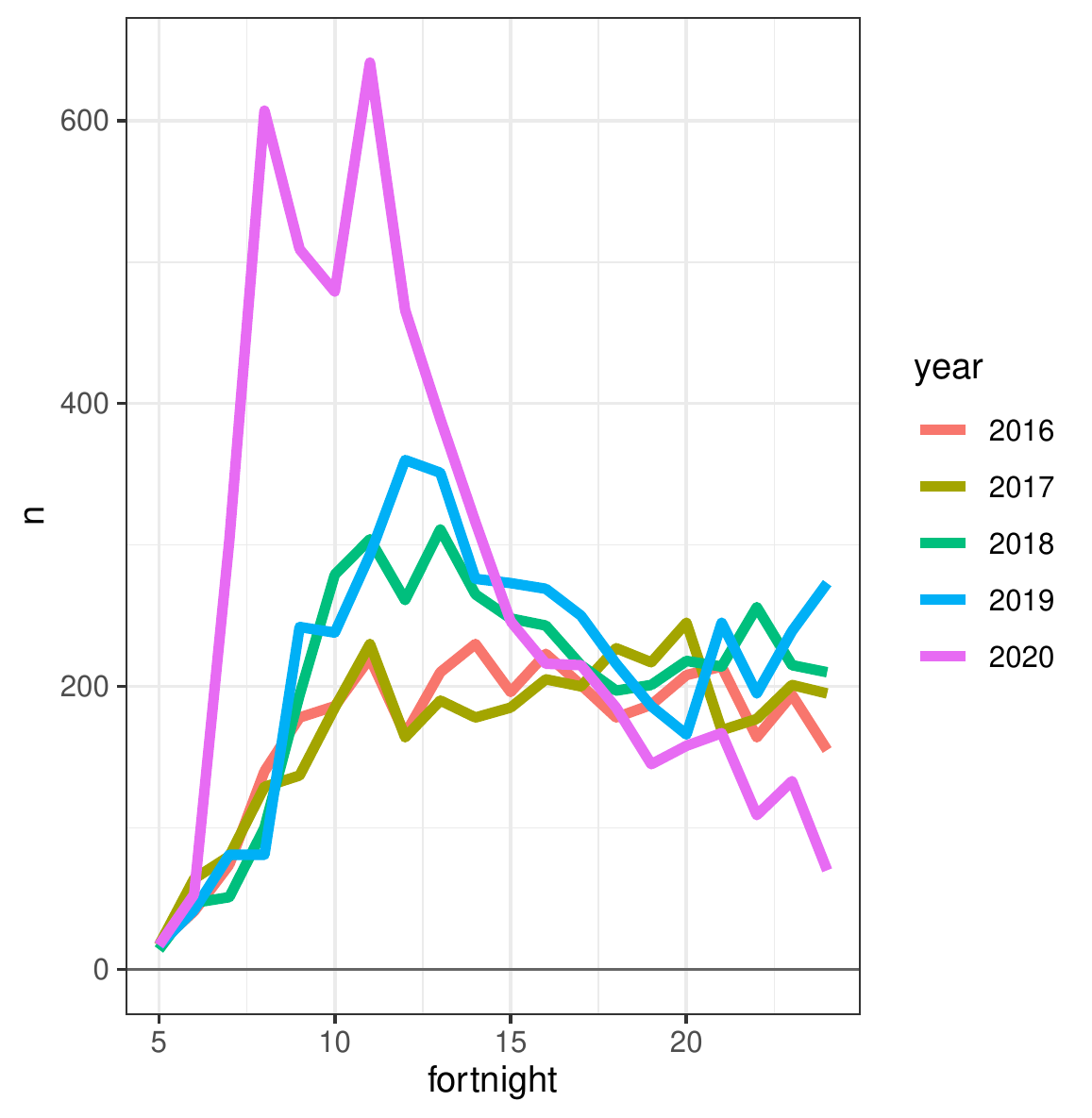}\label{fig:actA4}
	}
		\hfill
	\subfloat[Activity(5) $A$ - Optional]{
		\includegraphics[width=0.3\textwidth]{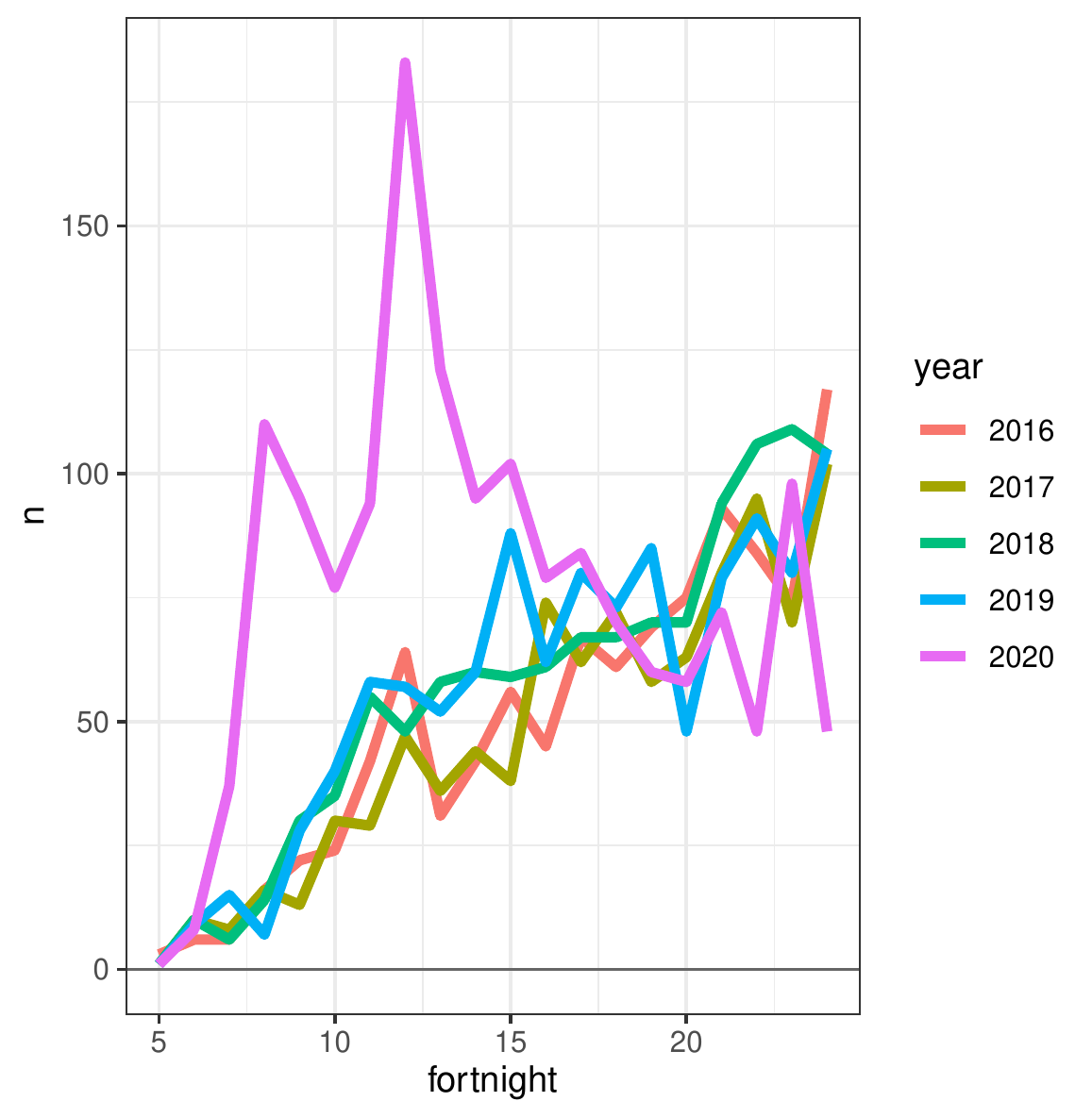}\label{fig:actA5}
		\hfill
	}
	\subfloat[Activity(6) $B$ - Optional]{
		\includegraphics[width=0.3\textwidth]{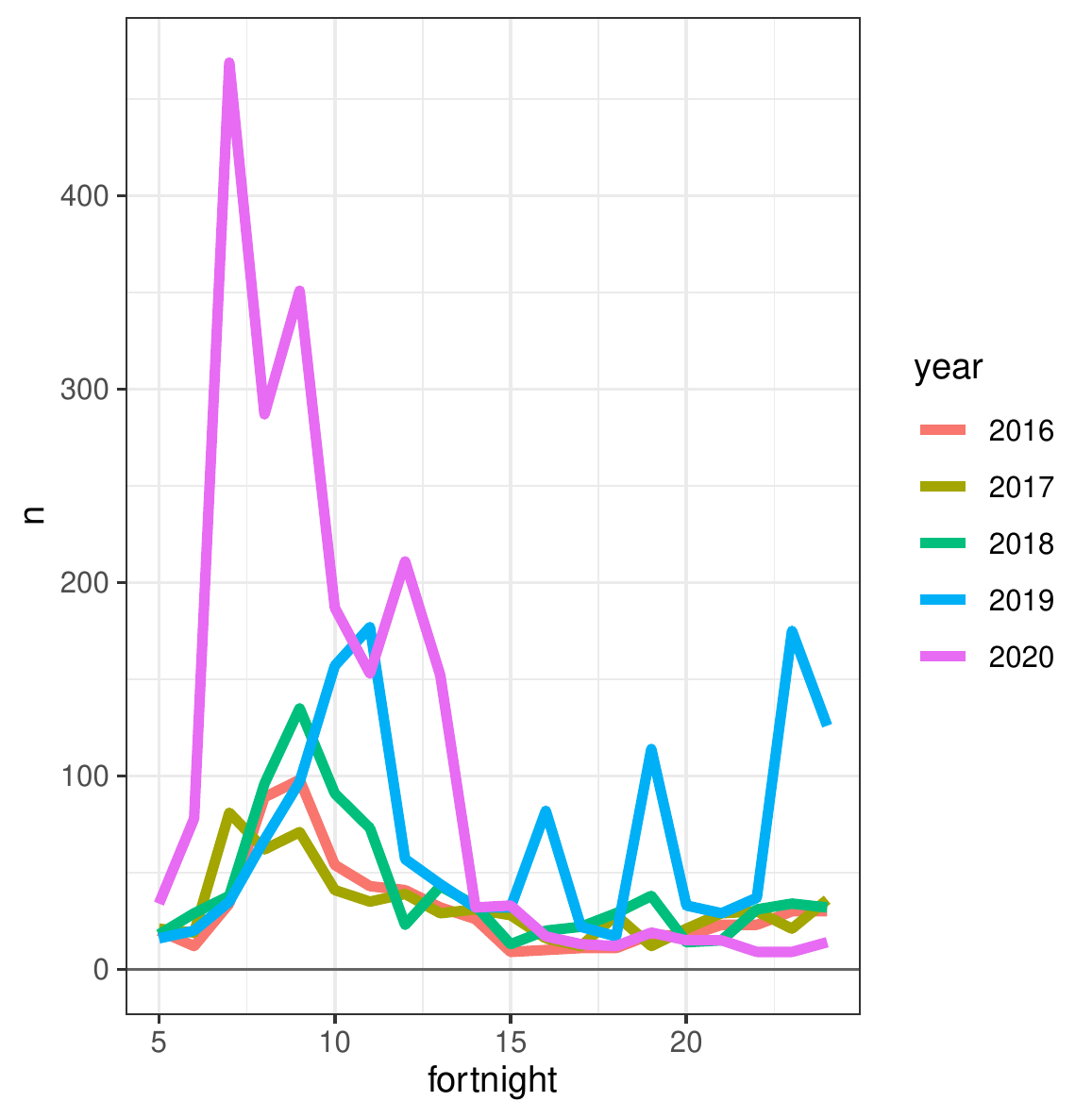}\label{fig:actA6}
	}
	\vspace{-4mm}
\caption{Vaccination process (most frequent behaviour only) -- observed activities over time (fortnightly), corresponding process models are captured in Figure~\ref{fig:vac-processes}}
\label{fig:vac-plots}
\end{figure}

Lastly, we analysed each of the process traces by looking into the distribution of the executed activities over time, we reported this information in Figures~\ref{fig:actA1} to~\ref{fig:actA6}. At this point, it is evident that the differences in behaviour were not only in terms of \emph{how} the activities were executed (i.e., their order and frequencies) but also \emph{when}. We summarised these findings in the following observation.

\begin{observation}\label{obs6}
In 2020, we can observe a clear (left-)shift (i.e., towards March) and early peak in the distribution of the activities executed within the most frequent behaviour of the vaccination process, as well as a different trend when compared to the past four years, which holds for all the activities involved in the vaccination process. Furthermore, the most frequent vaccination process in 2020 was more complex than the previous four years, allowing for more behavioural variants with frequently requiring additional activities (activity $A$, and $B$).
\end{observation}

The observed change can be explained by the recommendation that Australians receive their influenza vaccinations before the normal season (April-May). This recommendation was broadly advertised to minimize a possible double hit to the healthcare system: an epidemic of SARS-CoV-19 in addition to the usual Fall/Winter influenza season.~\footnote{https://www.abc.net.au/news/health/2020-04-01/australians-urged-to-get-flu-vaccination-coronavirus-covid-19/12107264} In the next section, we will discuss more in depth this observation from a medical perspective.

\begin{table}
\centering
{\scriptsize{
\bgroup
\def\arraystretch{1} 
\setlength{\tabcolsep}{5pt} 
  \caption{Top-20 traces, ordered by frequency}\label{tab:topbeh}
  \vspace{-4mm}
    \begin{tabular}{c|l|l|l|l|l}
    \hline
          & \textbf{2020} 
          & \textbf{2019}
          & \textbf{2018}
          & \textbf{2017}
          & \textbf{2016}\\\hline
          
    \textbf{1} & $\langle A \rangle$ 51894 & $\langle A \rangle$ 68277 & $\langle A \rangle$ 75049 & $\langle A \rangle$ 73148 & $\langle A \rangle$ 74130 \\
    \textbf{2} & $\langle A,G \rangle$ 15619 & $\langle A,A \rangle$ 18946 & $\langle A,A \rangle$ 21537 & $\langle A,A \rangle$ 22107 & $\langle A,A \rangle$ 22676 \\
    \textbf{3} & $\langle A,A \rangle$ 14154 & $\langle A,C,D \rangle$ 16172 & $\langle A,C,D \rangle$ 17090 & $\langle A,C,D \rangle$ 15889 & $\langle A,C,D \rangle$ 14487 \\
    \textbf{4} & $\langle A,C,D \rangle$ 12735 & $\langle A,B \rangle$ 13952 & $\langle A,B \rangle$ 13260 & $\langle A,B \rangle$ 13457 & $\langle A,B \rangle$ 13753 \\
    \textbf{5} & $\langle A,D \rangle$ 6367 & $\langle A,G \rangle$ 10885 & $\langle A,G \rangle$ 9145 & $\langle A,A,A \rangle$ 8994 & $\langle A,A,A \rangle$ 9267 \\
    \textbf{6} & $\langle A,B \rangle$ 5879 & $\langle A,A,A \rangle$ 7130 & $\langle A,A,A \rangle$ 8158 & $\langle A,G \rangle$ 6849 & $\langle A,G \rangle$ 6549 \\
    \textbf{7} & $\langle A,A,A \rangle$ 5492 & $\langle A,B,C,D \rangle$ 5013 & $\langle A,B,C,D \rangle$ 5154 & $\langle A,B,C,D \rangle$ 4497 & $\langle A,B,C,D \rangle$ 4646 \\
    \textbf{8} & $\langle A,G,A \rangle$ 3348 & $\langle A,C,D,A \rangle$ 3784 & $\langle A,E,F \rangle$ 4071 & $\langle A,A,A,A \rangle$ 4441 & $\langle A,A,A,A \rangle$ 4419 \\
    \textbf{9} & $\langle A,C,D,A \rangle$ 2850 & $\langle A,C,D,B \rangle$ 3539 & $\langle A,A,A,A \rangle$ 3950 & $\langle A,E,F \rangle$ 3947 & $\langle A,E,F \rangle$ 3968 \\
    \textbf{10} & $\langle A,A,A,A \rangle$ 2440 & $\langle A,E,F \rangle$ 3462 & $\langle A,C,D,A \rangle$ 3839 & $\langle A,C,D,A \rangle$ 3681 & $\langle A,C,D,A \rangle$ 3533 \\
    \textbf{11} & $\langle A,E,F,A \rangle$ 2256 & $\langle A,A,A,A \rangle$ 3231 & $\langle A,B,A \rangle$ 3174 & $\langle A,B,A \rangle$ 3331 & $\langle A,B,A \rangle$ 3469 \\
    \textbf{12} & $\langle A,C,D,B \rangle$ 2120 & $\langle A,B,A \rangle$ 2865 & $\langle A,E,F,A \rangle$ 3073 & $\langle A,E,F,A \rangle$ 2975 & $\langle A,E,F,A \rangle$ 2896 \\
    \textbf{13} & $\langle A,G,B \rangle$ 1893 & $\langle A,D \rangle$ 2774 & $\langle A,C,D,B \rangle$ 2592 & $\langle A,A,B \rangle$ 2553 & $\langle A,A,B \rangle$ 2660 \\
    \textbf{14} & $\langle A,E,F \rangle$ 1892 & $\langle A,G,A \rangle$ 2493 & $\langle A,G,A \rangle$ 2295 & $\langle A,A,A,A,A \rangle$ 2350 & $\langle A,A,A,A,A \rangle$ 2439 \\
    \textbf{15} & $\langle A,B,C,D \rangle$ 1880 & $\langle A,A,B \rangle$ 2406 & $\langle A,D \rangle$ 2118 & $\langle A,C,D,B \rangle$ 2275 & $\langle A,C \rangle$ 1809 \\
    \textbf{16} & $\langle A,A,C,D \rangle$ 1849 & $\langle A,E,F,A \rangle$ 2402 & $\langle A,A,B \rangle$ 1992 & $\langle A,C \rangle$ 1871 & $\langle A,G,A \rangle$ 1743 \\
    \textbf{17} & $\langle A,A,G \rangle$ 1578 & $\langle A,B,B \rangle$ 2156 & $\langle A,A,A,A,A \rangle$ 1932 & $\langle A,G,A \rangle$ 1791 & $\langle A,C,D,B \rangle$ 1720 \\
    \textbf{18} & $\langle A,A,D \rangle$ 1285 & $\langle A,A,C,D \rangle$ 1766 & $\langle A,A,C,D \rangle$ 1884 & $\langle A,A,C,D \rangle$ 1767 & $\langle A,B,B \rangle$ 1685 \\
    \textbf{19} & $\langle A,G,A,A \rangle$ 1267 & $\langle A,A,A,A,A \rangle$ 1555 & $\langle A,C \rangle$ 1879 & $\langle A,D \rangle$ 1682 & $\langle A,A,C,D \rangle$ 1654 \\
    \textbf{20} & $\langle A,A,A,A,A \rangle$ 1237 & $\langle A,A,G \rangle$ 1414 & $\langle A,B,B \rangle$ 1493 & $\langle A,B,B \rangle$ 1600 & $\langle A,D \rangle$ 1526 \\\hline
    \end{tabular}%
\egroup
}}
\end{table}%

\begin{table}
\centering
{\scriptsize{
\bgroup
\def\arraystretch{1} 
\setlength{\tabcolsep}{5pt} 
  \caption{Top-20 traces, extract of the traces including activity G.}\label{tab:topbehG}
  \vspace{-4mm}
    \begin{tabular}{c|l|l|l|l|l}
    \hline
          & \textbf{2020} 
          & \textbf{2019}
          & \textbf{2018}
          & \textbf{2017}
          & \textbf{2016}\\\hline
          
    \textbf{1} & $\langle A,G\rangle$ 15619 & $\langle A,G\rangle$ 10885 & $\langle A,G\rangle$ 9145 & $\langle A,G\rangle$ 6849 & $\langle A,G\rangle$ 6549 \\
    \textbf{2} & $\langle A,G,A\rangle$ 3348 & $\langle A,G,A\rangle$ 2493 & $\langle A,G,A\rangle$ 2295 & $\langle A,G,A\rangle$ 1791 & $\langle A,G,A\rangle$ 1743 \\
    \textbf{3} & $\langle A,G,B\rangle$ 1893 & $\langle A,A,G\rangle$ 1414 & -     & -     & - \\
    \textbf{4} & $\langle A,A,G\rangle$ 1578 & -     & -     & -     & - \\
    \textbf{5} & $\langle A,G,A,A\rangle$ 1267 & -     & -     & -     & - \\
\hline
    \end{tabular}%
\egroup
}}
\end{table}%

\subsection{From Process to Data Mining}\label{sec:challenge4}

Our process mining analysis provided us with a lead to follow. It allowed us to identify relevant differences in the behaviour of the patients in 2020, in particular, when considering the vaccination process. Accordingly, we refined our original research question into two sub-questions. \textbf{RQ1.} What type of vaccines have driven the frequency increase in the most frequent vaccination process traces? \textbf{RQ2.} What are the differences between vaccination behaviour of different age classes, i.e., children (0-17 years), adults (18-64 years), and elderly people (65+)?

To answer these research questions, process mining techniques can provide little help.
Process mining and, more in general, process science and process thinking can be a lighthouse in a ocean of data. However, it is difficult to dig deeper by only relying on process mining techniques, given that at the current stage they do not take into account \emph{rich} perspectives surrounding the process behaviour. In fact, to the best of our knowledge, there are no \emph{reliable} and \emph{effective} process mining techniques -- in the area of automated process discovery~\cite{augusto2018automated} and process variant analysis~\cite{taymouri2021business} -- that give a global picture of the process, taking into account \emph{all} the additional data recorded in the event attributes available in the event log.
For instance, to answer our research questions,
\begin{wraptable}[17]{r}{0.39\textwidth}
  \centering
  {\scriptsize{
  \caption{Encoding of vaccines}
  \vspace{-3mm}
    \begin{tabular}{r|l}
            \hline
            \textbf{Label} & \textbf{Provided Immunity}\\\hline
            \textbf{V1} & Cholera \\
            \textbf{V2} & Coxiella Burnetti \\
            \textbf{V3} & Diphtheria \\
            \textbf{V4} & Haemophilus B \\
            \textbf{V5} & Hepatitis A \\
            \textbf{V6} & Hepatitis B \\
            \textbf{v7} & HPV \\
            \textbf{V8} & Influenza \\
            \textbf{V9} & Japanese Encephalitis \\
            \textbf{V10} & Measles \\
            \textbf{V11} & Meningococcal \\
            \textbf{V13} & Pneumococcus \\
            \textbf{V14} & Poliomyelitis \\
            \textbf{V15} & Rabies \\
            \textbf{V16} & Rotavirus \\
            \textbf{V17} & Salmonella typhi \\
            \textbf{V18} & Tetanus \\
            \textbf{V19} & Tuberculosis \\
            \textbf{V20} & Varicella Zoster \\
            \textbf{V21} & Yellow Fever \\\hline
    \end{tabular}%
  \label{tab:vacmap}%
    }}
\end{wraptable}%
the crucial event attributes are \emph{patient age} and \emph{vaccine type}, but the integration of this information in a process model is not a trivial problem. Besides, we note that our refined research questions are data mining oriented. In fact, process mining and data mining are complimentary, and future research directions should leverage this relation between the two disciplines to bring them together. 

To continue with our analysis, we extracted \emph{all} the data regarding vaccination events (activity $G$) from the original dataset (GP16-20 event log), and we analysed the different vaccine types and the immunity they provide, Table~\ref{tab:vacmap} shows a mapping between the vaccines and the labels we will use to simplify the presentation of the data.

\begin{figure}[H]
	\centering
	\subfloat[Vaccinations count in 2020, by age class]{
		\includegraphics[width=0.97\textwidth]{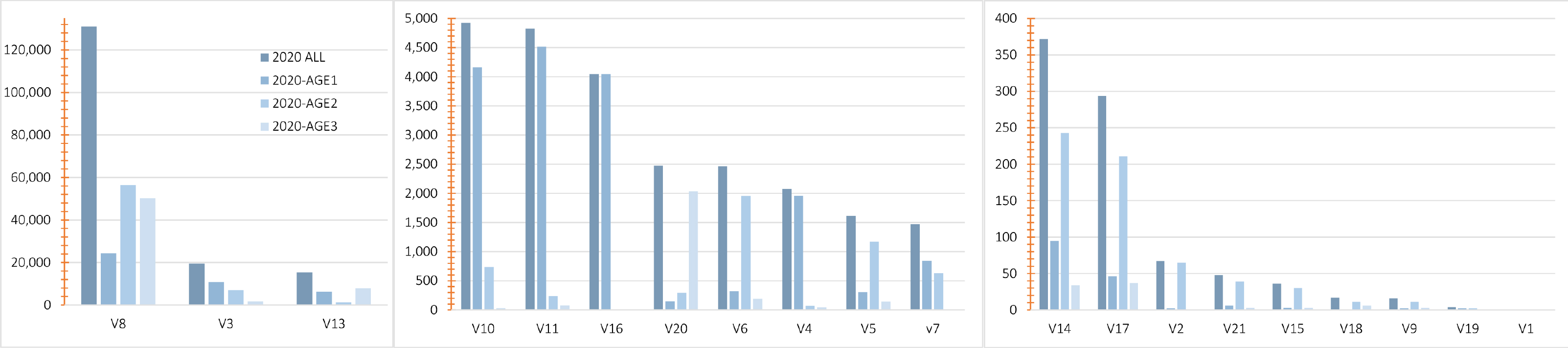}\label{fig:vacount}
	}
		\\
	\subfloat[Change in vaccination count -- all ages (2020 is the reference year) ]{
		\includegraphics[width=0.97\textwidth]{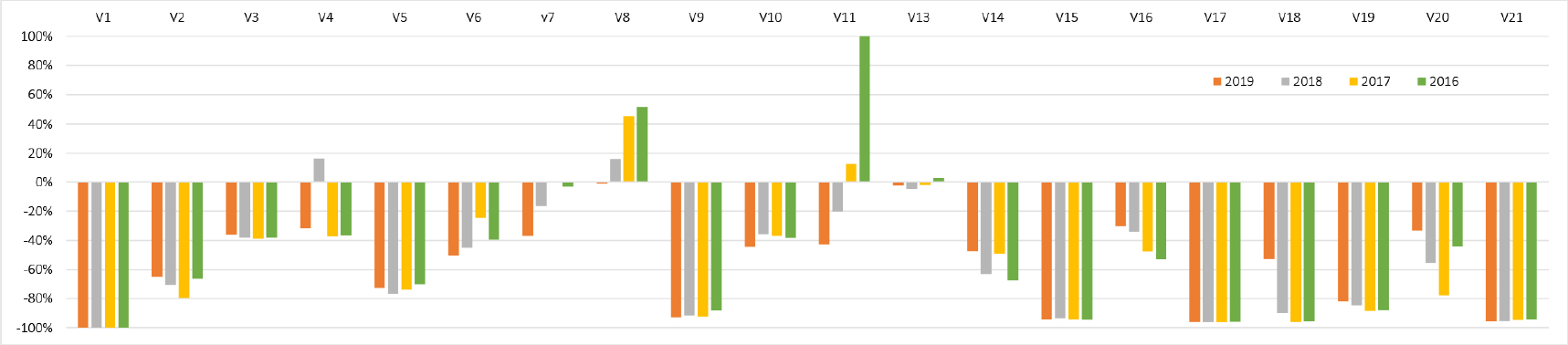}\label{fig:vachange0}
	}
		\\
	\subfloat[Change in vaccination count -- 0-17 years old (2020 is the reference year) ]{
		\includegraphics[width=0.97\textwidth]{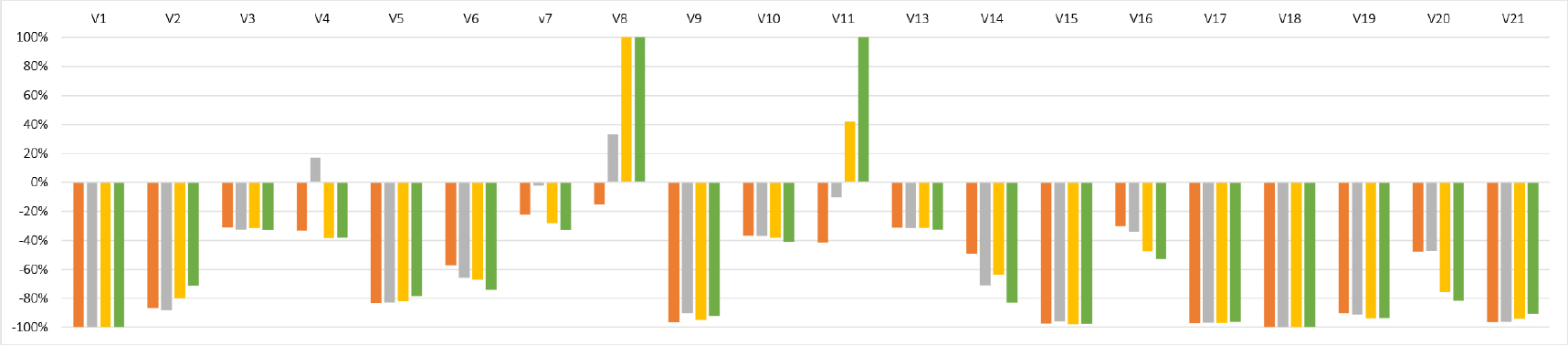}\label{fig:vachange1}
	}
		\\
	\subfloat[Change in vaccination count -- 18-64 years old (2020 is the reference year) ]{
		\includegraphics[width=0.97\textwidth]{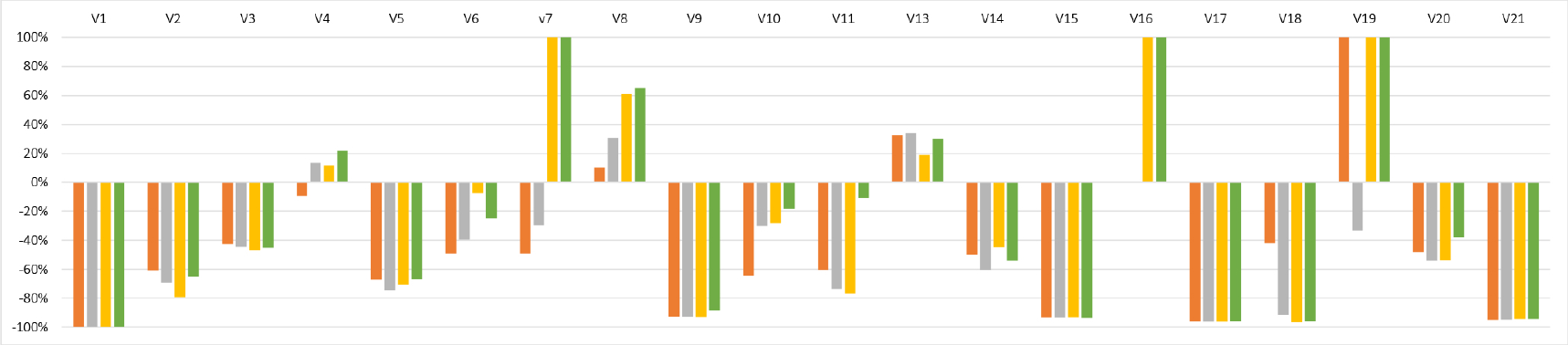}\label{fig:vachange2}
	}
		\\
	\subfloat[Change in vaccination count -- 65+ years old (2020 is the reference year) ]{
		\includegraphics[width=0.97\textwidth]{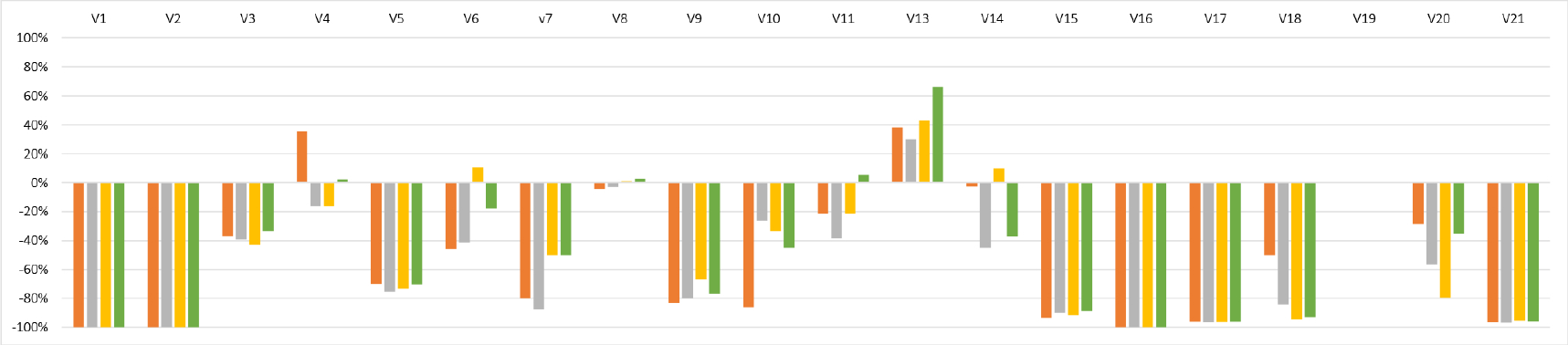}\label{fig:vachange3}
	}
	\vspace{-4mm}
\caption{Vaccinations comparison, years 2020-2016}
\label{fig:vac-plots}
\end{figure}

Figure~\ref{fig:vacount} shows the absolute number of vaccines we observed in 2020, grouped by the provided immunisation (see Table~\ref{tab:vacmap}). Figure~\ref{fig:vachange0} to~\ref{fig:vachange3} report the change in the absolute number of vaccines observed in 2020, when compared to the past four years. We compared the vaccination count by grouping the patients by age, specifically: \emph{all ages} (Figure~\ref{fig:vachange0}); young people (0 to 17 years old -- Figure~\ref{fig:vachange1}); adults (18 to 64 years old -- Figure~\ref{fig:vachange2}); elderly people (65+ years old -- Figure~\ref{fig:vachange3}). From the data, we can draw the following observation.

\begin{observation}\label{obs7}
In 2020, there was a surge of \emph{influenza} (V8) and \emph{pneumococcus} (V13) vaccinations (see Figure~\ref{fig:vac-plots}, vaccine V8 and V13), predominant in adults and elderly people, and in contrast with a decrease of these vaccinations for young people (see Figures~\ref{fig:vachange0} to~\ref{fig:vachange3}, V8 and V13). The increase is even more startling when we consider that all the other vaccines suffered a decrease of approximately 50\% (on average).  
\end{observation}

Similar to our discussion on Observation~\ref{obs6}, the surge in influenza vaccinations can be linked to the public health campaign aiming at increasing the proportion of patients receiving the influenza vaccine to reduce the size of the seasonal peak of influenza infections and hospital admissions, in anticipation of the potential overload of the healthcare system by COVID-19 patients. Similarly, a larger proportion of older adults might have received their pneumococcal vaccines concomitantly. Furthermore, we were able to observe that vaccines associated with international travel requirements (i.e. Yellow Fever, Japanese Encephalitis) practically disappeared, probably a result of international border closures. 

In the next section, we explore more in depth these implications from a medical point of view.

\subsection{Limitations of the Study}

In this section, we described how we have analysed the data and the observations we could draw from it. 

To analyse the data, we followed a well-known methodology, PM\textsuperscript{2}~\cite{van2015pm}. We note that the latter was designed to be applied in a business-context. However, we argue that this does not pose a threat to its applicability in healthcare, in fact, we were able to adhere to its stages from start to end, with the exception of omitting the execution of the \emph{process improvement} stage, since it was out of the scope of this study. 

To execute the process mining analysis, we relied on a subset of the existing state-of-the-art process mining techniques for automated process discovery and variant analysis, which we selected according to the findings of the most recent literature reviews. While, in theory, applying other techniques may have yielded different or better results, we recall that the process mining techniques we used were the latest and the most reliable.

The observations reported in this study cannot be generalised to Australia, nor the state of Victoria. However, we note that the analysed data captured the behaviour of approximately 400 thousand patients (per year), which account for almost 6\% of the entire population of the state of Victoria -- a remarkable percentage, especially when we consider that not the whole population regularly visit GP clinics. While the observations reported in this study are derived from the data and, hence, objective, their analysis and our discussion represent our interpretation. We note that when providing an explanation for a specific observation we considered findings of other similar studies and the experience of two domain experts who co-authored this study (Dr Capurro and Dr Manski-Nankervis). In theory, alternative interpretations for some of our observations may be possible but, to the best of our knowledge, the one we provided in this study are the most reasonable and realistic.

\section{Discussion}\label{sec:discussion}

The study presented here represents the first use of process mining techniques to analyze the impact of the COVID-19 pandemic in health services utilization patterns in primary care. Using a combination of process mining techniques we were able to highlight several relevant changes in health services utilization patterns associated with the disruptions seen in 2020. In addition to these, we were able to highlight some limitations of the process mining tools available today, in particular, when applying them to analyze healthcare process data.

Overall, we observed a widespread reduction of GP activities during the period included in our study, when compared to the same period in the four preceding years, concordant with what has been reported in other countries~\cite{baum2021reduced}. It is expected that such a reduction of GP activities led to a reduction of specialists visits --given that the Australian is a referral-based system. The consequences of such additional potential reduction of healthcare activities remain to be seen. From the process perspective, in such a situation, we would have expected a reduction in the number of distinct healthcare process execution, instead, the degree of variety of process behaviour remained almost unchanged during this period. 

One activity that showed a different behavior were drug prescriptions. We observed an increase in drug refill prescriptions, with peaks in March, April, July and September. These peaks are associated with periods immediately before lock-downs and might represent overstocking of chronic medications. This observation is in line with what has been observed in Australian national drug prescription databases~\cite{mian2021increased}.

The most notable changes were observed in activities involving vaccinations. First, we see that although there still was a reduction in the total number of vaccinations, the drop was relatively minor compared to the rest of the GP activities. Vaccinations dropped an average of 1.3\% and all other activities dropped an average of 23.6\%. This contrasts to what has been reported elsewhere, where the 2020 pandemic has been associated to significant reduction in vaccination rates~\cite{Gaythorpe2021}. When we look into specific vaccines, we can see an increase in influenza vaccinations together with an earlier peak. This is in line with public health campaigns urging citizens to get their annual influenza vaccines and prevent a double epidemic. Interestingly, in older adults we can see a parallel increase in pneumococcal vaccinations. The most likely explanation was the drive to reduce any preventable respiratory infection in preparation of the impending pandemic. Finally, vaccines normally recommended for international travel (Yellow Fever, Japanese Encephalitis, Cholera) practically disappeared, as a consequence of the severe limitations to international travel.

From the process mining perspective we faced several challenges related to the problem of analysing a vast amount of process execution data. To the best of our knowledge, the event log analysed in this study represents the largest real-life event log used for automated process discovery and process variant analysis, especially, in the healthcare context. We showed that traditional process mining tools present some limitations when attempting to analyze processes with high behavioural variability.

The first challenge consisted of imprecise timestamps, since the time granularity was limited to day-level, a recurrent problem in the healthcare context that yet has to be solved. In our case, we relied on clinical knowledge to address this issue by defining a sequence of clinically meaningful activities as a tie-breaker for activities that had identical timestamps. 

The second challenge involved the identification of start and end events for a process that is, by nature, unbounded. Once again, we relied on domain expertise to overcome this problem and we presented a generalisation of our solution, suitable for various contexts, in the algorithm described in Section~\ref{sec:analysis}.

The third challenge was the amount of data itself, and its high behavioral variability, which disarmed state-of-the-art process mining techniques for automated process discovery and process variant analysis. Although the scope of this study was not to devise novel variants of these techniques to deal with such type of data, we highlighted possible directions for future research addressing the improvement of these techniques.

Lastly, our study highlights that process mining techniques cannot yet leverage the event log information that is not related to the process behaviour and control-flow (e.g., patient age, medications, etc). This requires process analysts to integrate process mining analysis with data mining analysis. While this problem could be solved straightforwardly by further analysing the data from a different perspective, we call for future analysis methodologies and tools that automatically integrate both process and data perspectives.

\section{Conclusion}\label{sec:conclusion}

This study represents the first application of process mining techniques to analyze the impacts of the COVID-19 pandemic in the patterns of primary care service utilization, specifically, in the General Practice day-to-day  healthcare processes of Victorian~\footnote{Victoria, Australia} patients. Our analysis identified several relevant changes in the behavioural patterns of the patients. While some of these changes were expected, i.e., overall reduction in number of attended GP visits, some were not, i.e., increase in the number of medication prescriptions, less than expected drop in vaccinations, and increase of influenza and pneumococcus vaccinations -- in contrast with research findings from different geographical areas~\cite{santoli2020effects,Gaythorpe2021,lassi2021impact}.

The size of the data-set under analysis -- counting 31-million events -- and the variability of the observed process behavior were unique, and the challenges we faced and overcame during the process mining analysis clearly highlighted the need for improving existing process mining techniques, drawing directions for future work. In particular, future process discovery techniques should integrate in the discovered process models also data surrounding the process behaviour and its control flow. In the healthcare context, this data is the information capturing a patient profile (e.g., age, gender, etc) and their medical procedure (e.g., type of vaccination or prescribed medication). Furthermore, existing process mining techniques are not tailored to deal with large amount of data that captures highly variable behaviour. Future research should consider the design of methods that can \emph{automatically} filter process execution data to detect and extract the most relevant/interesting process behaviour (not necessarily the most frequent) by analysing the outputs of a set of process mining techniques (e.g., a combination of automated process discovery and process variant analysis). Lastly, as process mining applicability in the healthcare context gains momentum, novel process mining techniques should be tailored for such a context and leverage domain expertise to increase their effectiveness.

\bibliography{lit}

\end{document}